\documentclass[a4paper,fleqn]{cas-sc}
\usepackage[authoryear]{natbib}
\usepackage[font=it,skip=3pt]{subcaption}
\graphicspath{{.}}

\makeatletter
\g@addto@macro\UrlBreaks{\do\-}
\makeatother

\renewcommand{\it}[1]{\textit{#1}}
\newcommand{\mt}[1]{\texttt{\detokenize{#1}}}

\newcolumntype{X}[1]{>{\raggedright\arraybackslash}m{#1\linewidth}}
\newcolumntype{Y}[1]{>{\centering\arraybackslash}m{#1\linewidth}}
\newcolumntype{Z}[1]{>{\raggedright\arraybackslash}p{#1\linewidth}}

\begin{document}

\let\WriteBookmarks\relax
\def\floatpagepagefraction{1}
\def\textpagefraction{.001}

\shorttitle{Data-Driven Mixed-Methods Framework for Cybernetic Urban Mobility Governance}
\shortauthors{O. Yusuf et~al.}
\title [mode = title]{Data-Driven Mixed-Methods Framework for Cybernetic Urban Mobility Governance}

\author[1]{Oluwaleke Yusuf}[orcid=0000-0002-5904-648X]
\ead[url]{https://www.ntnu.edu/employees/oluwaleke.u.yusuf}

\author[1]{Morten Breivik}[orcid=0000-0002-0457-1850]
\ead[url]{https://www.ntnu.edu/employees/morten.breivik}

\author[1]{Adil Rasheed}[orcid=0000-0003-2690-983X]
\ead[url]{https://www.ntnu.edu/employees/adil.rasheed}

\affiliation[1]{
    organization={Department of Engineering Cybernetics, Norwegian University of Science and Technology (NTNU)},
    city={Trondheim},
    citysep={},
    postcode={NO-7491},
    country={Norway}
}

\begin{abstract}[S U M M A R Y]
    This study develops a cybernetically inspired mixed-methods framework that bridges the gap between policy formation and implementation through feedback-driven analysis of mobility transitions. Using a major campus consolidation in Trondheim, Norway as a case study, we examine how this framework supports sustainable mobility through integrated analysis of mobility patterns, constraints, and transition impacts. The consolidation eliminates over 1,300 parking spaces while increasing daily population by 9,300 people.
    We employ a mixed-methods approach combining qualitative survey data (n=573) with quantitative big data analysis of public transit and crowd movement patterns. This integrates three analytical components and provides grounded insights into commuting flows, modes, durations, distances, and congestion points, while addressing the spatiotemporal mobility realities of affected populations.
    The analysis reveals complex mobility constraints, with 59.3\% of respondents having children and private cars dominating in winter (49.4\%). Though there is broad support for sustainable mobility goals, 86.0\% identify increasing travel duration as primary difficulty. Quantitative analysis highlights peak usage patterns and congestion risks, with seasonal variations.
    This study demonstrates integrating qualitative and quantitative analysis to anticipate negative impacts and enable efficient sustainable mobility policies. The results inform practical recommendations for data-driven mobility interventions that align sustainability goals with lived realities.
\end{abstract}


\maketitle

\section{Introduction}
\label{sec:introduction}
Sustainable mobility policy initiatives often fail to achieve their desired objectives due to insufficient analytical capabilities that can bridge the gap between policy formation and implementation \citep{Wikstrom2024smt,Muller_Eie2023smi}. Current approaches frequently rely on simplified assumptions about mobility behaviour, inadequately accounting for the spatiotemporal realities and individual constraints that shape daily mobility choices \citep{Whitmarsh2012hui,OliveiraSoares2025biu}. The challenge lies in developing sophisticated technical solutions that can integrate diverse data sources to understand mobility transitions. Enhanced mixed-methods analytical frameworks are needed that can capture both measurable patterns and subjective experiences, while revealing opportunities for adaptive governance approaches \citep{Yusuf2025lbd,Banister2008tsm}.

The Norwegian University of Science and Technology (NTNU) provides a compelling prototype for examining these systemic governance challenges in practice. The university is undergoing major campus consolidation at Gløshaugen, which drastically reduces parking availability \citep{NTNU2023cc,Bjerva2025dps}. While the study focuses on the immediate behavioural and perceptual effects of NTNU's parking policy, it also exposes the need for a systemic governance perspective capable of learning from feedback.

The Hesthagen parking lot, previously a key location for employees and students, has closed permanently on March 1, 2025 \citep{Javorovic2025spm}, to make way for the new Economics and Innovation building. Although construction has been slightly delayed, the parking area will remain closed. The project, part of the university's broader Campus Collection initiative, aims to bring departments such as Industrial Economics and Technology Management into a central cluster with the existing Adolf Øien building \citep{NTNU2023cc}. As a bounded socio-technical system, NTNU's transformation presents both opportunities and challenges for implementing cybernetic governance principles.

The consolidation presents challenges for commuters by car, with NTNU's overall parking coverage shrinking from six percent to under two percent per person \citep{Maele2024nrp}. The most central and affordable parking areas are being eliminated or reserved for those with special permits, such as those with mobility issues \citep{Bjerva2025dps,Maele2024nrp}. The current policy offers free parking for electric vehicles (except in select zones), while public transit has different cost structures. For example, a monthly bus pass costs more than parking fees, which can influence the choice of travel mode \citep{Maele2024nrp}.

To counterbalance the removal of at least 1,314 car spaces across Gløshaugen, Moholt, and Dragvoll \citep{Bjerva2025dps}, NTNU is increasing bicycle and public transit infrastructure. New indoor bicycle parking facilities are being constructed, with 2,462 spaces planned across six projects \citep{Bjerva2025dps}. A minimum quota ensures adequate bike capacity per student and employee. Moreover, NTNU is in talks with transportation providers to improve bus and train services, anticipating greater dependency on these options as more students and employees concentrate at Gløshaugen \citep{Bjerva2025dps}.

NTNU is also revisiting its parking regulations, which have been in place since 2014. These include paid parking for fossil fuel vehicles and prioritise revenue use to support sustainable mobility methods \citep{Maele2024nrp}. However, gaps in the local permit system have been identified, especially regarding the vague criteria for health-based and childcare-related parking needs. A new regulatory framework is expected in 2024, aiming to protect vulnerable groups while aligning with Trondheim's environmental and urban development policies \citep{Maele2024nrp}.

\vspace{\baselineskip}
The increased population growth in urban areas has placed increased strain on mobility systems and their capability to adequately serve the needs of residents. With urban sizes and population expected to increase at accelerated paces \citep{Gulc2024cos}, the sustainability of contemporary mobility solutions and their impact on urban quality of life have been frequently discussed \citep{Ritchie2020cpt}. There is now a shift in mobility policies away from car-centred to those with a people-centric focus \citep{Millard_Ball2011awr,Jones2016teo}, prioritising various forms of public transit and active mobility over private car use. However, the complexity of mobility transitions requires more than modal shift strategies.

The multi-level perspective in transport studies emphasises that systemic transitions entail co-evolution and multi-dimensional interactions between industry, technology, markets, policy, culture and civil society \citep{Geels2012ast}. This systems thinking approach recognises that effective mobility interventions must address not only infrastructure and technology but also the social practices and governance mechanisms that shape mobility behaviour \citep{Whitmarsh2012hui}. Research on multilevel policy mixes demonstrates that sustainable urban mobility transitions require complex coordination across governance levels, with national funding biases and limited cross-level cooperation often hampering progress \citep{Liu2024imp}.

Studies of spatial accessibility show that local and regional factors jointly influence travel behaviour, requiring integrated approaches that consider both walkability and broader transport connectivity \citep{Lussier_Tomaszewski2021tra}. NTNU's campus consolidation at Gløshaugen exemplifies these broader systemic challenges, with relocating the humanities and social sciences faculty from Dragvoll, along with the health and education faculties from Øya and Kalvskinnet, resulting in a net increase of 8,000 students and 1,300 employees based at Gløshaugen \citep{Ramboll2024kfv} in around 90,000 square metres of new buildings. By the planned completion in 2030, the campus development unit estimates that nearly 40,000 students and 7,000 employees will be based at the central campus \citep{NTNU2025ocn}.

The practical constraints on available building space make the removal of parking spaces appear to be a logical step, leading to the elimination of $\approx 80\%$ of available parking spaces without replacement. This scale of transformation of mobility infrastructure presents both opportunities and challenges for the implementation of sustainable mobility policy. Any issues stemming from such extensive changes today are likely to be magnified by the increased demands and pressure on mobility systems as the campus consolidation progresses and the daily population grows by approximately 9,300 students and employees. Furthermore, studies have shown that transitions towards sustainable mobility require major changes in culture and institutional support \citep{Kohler2020lct}.

The broader challenge facing such mobility transitions aligns with research showing that current initiatives to increase the adoption and share of sustainable mobility modes, not fully accounting for the realities of daily mobility and individual spatiotemporal constraints, often fail to achieve their desired objectives \citep{Wikstrom2024smt,Muller_Eie2023smi,Camilleri2022ubt}. This, despite research showing that said sustainable modes result in relatively higher satisfaction and positive emotions among users \citep{Mouratidis2023stm}, while reducing the share of car-based trips \citep{Remme2022wbf}. Research on transport poverty demonstrates that households caught in sustainability transitions may face exclusion from essential activities, particularly when policy measures fail to account for diverse mobility needs and constraints \citep{DallaLonga2025sit}.

This challenge highlights the importance of context-sensitive approaches that can bridge the gap between policy intentions and implementation realities. Studies have emphasised that effective sustainable mobility solutions must be aligned with the specific needs of the populations and the spatial characteristics of each location \citep{Wismadi2025sat,Borowska_Stefanska2025pot}. Research on citizen involvement demonstrates the value of multi-level participation frameworks that connect strategic and tactical decision-making in mobility planning \citep{Ibeas2011cii}. Studies of public transit competitiveness show the importance of geospatial analysis in identifying spatial imbalances and vulnerable areas requiring targeted policy interventions \citep{Hwang2025esu}.

More critically, \citet{Yusuf2025lbd} identified the potential of emerging technologies, such as digital twins, big data analytics, and artificial intelligence, to enhance analytical capabilities for more effective policy interventions that can anticipate and address the diverse needs of affected populations. Research on campus transportation strategies emphasises the importance of decision-making frameworks that account for immediate feasibility, cost-effectiveness, and long-term sustainability implications \citep{Bakioglu2024sos}.

\vspace{\baselineskip}
The mobility policy transitions associated with campus consolidation have generated considerable public discourse and commentary, most visible on the Universitetsavisa platform \citep{Bjerva2025dps,Maele2024nrp}. Among the 71 reactions recorded in both articles, the responses were predominantly negative, expressing sadness and anger. The comments sections revealed diverse perspectives on the implementation of mobility policies, some highlighting concerns about how mobility costs are presented, and others noting that public transit may present challenges for those with multiple commitments, such as caregiving responsibilities or lengthy commutes. In a separate opinion piece \citep{Ansatte2025npm}, several employees expressed concern that changes in mobility policy could affect work-life balance, recruitment, and productivity.

They questioned the rationale behind certain policy decisions, noting that many already use electric vehicles. The authors suggested a more inclusive strategy that supports sustainability goals while addressing the practical needs of campus employees. The broader concern reflected in these discussions highlights the challenge of implementing mobility policies that account for various life situations. While some advocate for reduced car dependency in central campuses, others point to logistical challenges faced by families, disabled individuals, and those commuting from areas with limited public transit connectivity.

There is also concern about infrastructure gaps, such as missing modal connections and cycling infrastructure. Research on effective policy packages emphasises that successful interventions must address all elements of practice as well as wider networks of practices, requiring enabling measures beyond traditional push and pull approaches \citep{Albers2025dep}. This discourse suggests a need for more sophisticated approaches to the development of mobility policy that can bridge the gap between sustainability goals and lived realities. For effective policy implementation, there is a clear need for improved analytical capabilities that can anticipate impacts and inform data-driven decision making.

\vspace{\baselineskip}
To address the aforementioned challenges in sustainable mobility policy development, this study demonstrates how sophisticated technical solutions can bridge the gap between policy formation and implementation. Using NTNU's campus consolidation at Gløshaugen as a case study, we examine how enhanced analytical capabilities can support data-driven decision making and stakeholder engagement in mobility transitions. The investigation employs a mixed-methods approach combining qualitative survey data with quantitative big mobility data analysis to understand spatiotemporal mobility realities and needs of affected populations.

The qualitative component involves an anonymised survey collecting data on commuting patterns, mobility preferences, daily schedules, and the impacts of mobility transition on choices, family logistics, and personal expenses. It examines barriers respondents may encounter when shifting to alternative mobility modes, including connectivity issues, distance to public transit, and infrastructure adequacy. The quantitative component leverages big mobility data to analyse passenger dynamics and people flows at bus stops and roads surrounding Gløshaugen.

Data on daily schedules, such as start and end times and the on/off-campus work ratio, enable querying of available datasets regarding current mobility dynamics and potential changes as campus consolidation leads to increased travel volumes and shifting spatiotemporal movement patterns. Furthermore, respondents' accounts of their daily mobility needs and suggestions for practical measures offer valuable insights into individual experiences that may otherwise be overlooked in broader policy analyses.

The combination of these approaches demonstrates how mixed-methods analysis can anticipate negative impacts and enable more efficient, proactive policies toward sustainable mobility futures, ultimately informing strategies that align sustainability goals with lived realities. This study also recognises that mobility policies are not isolated interventions but elements within complex socio-technical systems. Decisions about parking or commuting influence and are influenced by broader feedback loops connecting individual behaviour, institutional design, and urban infrastructure. To capture this systemic interplay, we adopt a mixed-methods framework that not only measures outcomes but also interprets them as signals within a larger governance architecture.

\vspace{\baselineskip}
The primary contributions of this study can be summarised as follows:

\begin{enumerate}
    \item \textbf{Mixed-Methods Framework for Mobility Policy Analysis}: We demonstrate a comprehensive analytical framework that integrates qualitative survey data (n=573) with quantitative big mobility data to reveal critical gaps between policy intentions and implementation realities, providing grounded insights into spatiotemporal mobility patterns and constraints.
    \item \textbf{Enhanced Analytical Capabilities for Policy Development}: Through the NTNU case study, we provide empirical evidence of mobility transition impacts, revealing significant seasonal variations, distance-dependent mode competitiveness, and specific congestion bottlenecks that inform infrastructure planning and policy design for campus consolidation scenarios.
    \item \textbf{Data-Driven Policy Recommendations}: We deliver targeted interventions that address both immediate challenges and long-term sustainability goals, demonstrating how mixed-methods analysis can anticipate negative impacts and guide proactive policy development before implementation.
    \item \textbf{Transferable Methodological Approach}: We establish a replicable methodology that combines big data analytics with stakeholder feedback, enabling policymakers to simulate interventions and assess impacts across diverse urban contexts. The framework also demonstrates how analytical insights can inform adaptive governance approaches that treat policy outcomes as feedback for continuous learning and system improvement.
\end{enumerate}

\vspace{\baselineskip}
The rest of this paper is organised as follows:
\autoref{sec:methodology} details the survey design, data collection procedures, and analytical approach, including both quantitative and qualitative methods for the mixed-methods framework.
\autoref{sec:results-discussion} presents the main findings on mobility patterns and transition impacts, discussing implications for sustainable mobility policy. The discussion synthesises the insights of both the qualitative survey \autoref{sec:results-qualitative} and the quantitative analysis of the big mobility data \autoref{sec:results-quantitative}, as well as the integrated mobility analysis in \autoref{sec:results-mobility}.
\autoref{sec:cybernetics} provides a cybernetic perspective on the findings, demonstrating how the NTNU case exemplifies broader challenges in governance architecture and proposing adaptive feedback frameworks for policy learning.
Finally, \autoref{sec:conclusion} summarises the key results, highlights the study's contributions and limitations, and outlines recommendations for future research and policy development. The potential for cybernetic governance approaches in urban mobility is discussed, along with practical recommendations for various stakeholders.

\section{Methodology}
\label{sec:methodology}

\subsection{Qualitative: Anonymous Survey}
\label{sec:methodology-qualitative}
Prior to the distribution of the survey, approval was obtained from the Norwegian Agency for Shared Services in Education and Research (Sikt) for the processing of limited personal data (age group, gender, role, commuting patterns and family circumstances), which is essential to understand the various commuting needs and constraints across different groups at NTNU. See attached PDF \textit{``Sikt: Assessment of processing of personal data – Ref. 874959''} in the Appendix for details.
The survey was hosted on the Nettskjema platform and promoted via NTNU's Innsida channel, which is accessible to all students and employees \citep{Danielsen2025hbd}. See attached PDF \textit{``Innsida: How are you affected by the reduction in the number of parking spaces?''} in the Appendix for details.

In line with NTNU practices and to reach a broad audience, the survey was provided in both Norwegian and English, and remained open from 1st to 11th April 2025. The questionnaire, identical in both versions, was organised into five thematic sections. See the attached PDF \textit{``Nettskjema: Survey on Mobility and Commuting Patterns at NTNU''} in the appendix for a detailed overview of each section. In summary, the themes are as follows:

\begin{enumerate} \itemsep=0pt
    \item \it{General Information} which gathers basic demographic data such as gender, age group, and the respondent's position at NTNU.
    \item \it{Commute and Transport} which focuses on commuting frequency, seasonal mobility mode choices, and proximity to public transit stops.
    \item \it{Family and Daily Schedule} where respondents indicate whether they have children, specify their working hours, and describe their remote versus on-site work balance.
    \item \it{Impact of Parking Reduction} that examines anticipated changes in commute duration, household mobility costs, and flexibility in adopting alternative mobility modes.
    \item \it{Impacts and Recommendations} through which participants reflect on potential job implications, identify main obstacles to alternative commuting, and suggest practical measures that NTNU could implement.
\end{enumerate}

At the end of the survey period, a total of 574 responses were received: 471 from the Norwegian version and 103 from the English version. Following initial data cleaning and preprocessing, the final dataset comprised 573 rows (responses) and 69 columns (questions). The preprocessing involved:

\begin{itemize} \itemsep=0pt
    \item Merging both versions into a unified English dataset and translating all free-text and open-ended responses to English.
    \item Semantic deduplication of free-text responses, i.e. user-typed inputs where no suitable drop-down option was selected, into unified labels.
    \item Manual corrections and refactoring of selected responses to reduce noise and simplify analysis.
    \item Engineering stop coordinates and parent municipalities based on data obtained from Entur for Trøndelag (AtB) \citep{Entur2025sat}.
\end{itemize}

\begin{figure}[tb!]
    \centering \includegraphics[width=.99\textwidth]{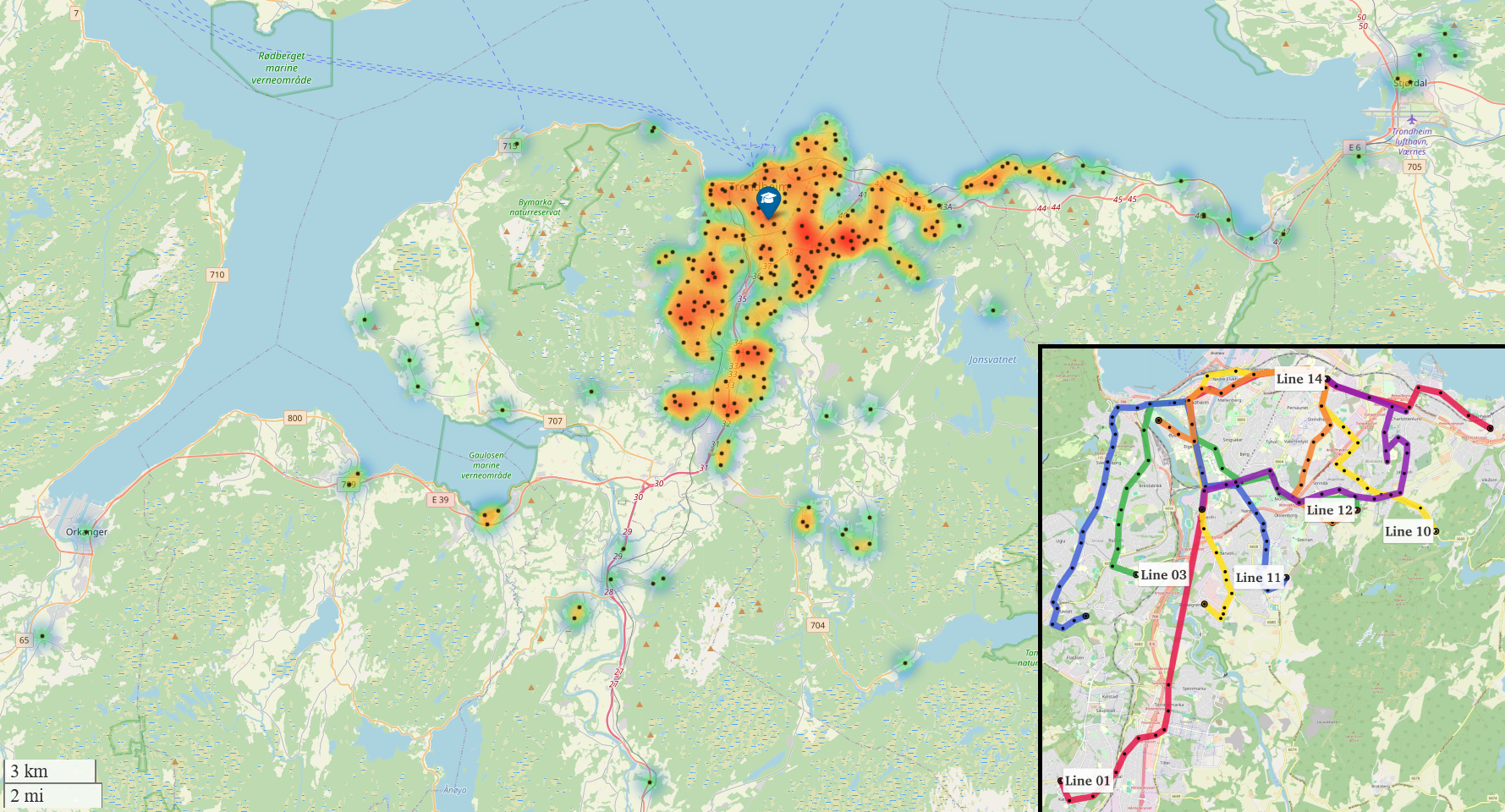}
    \caption{Heatmap depicting the geographical distribution of bus stops nearest to the home of survey respondents. The inset map shows the bus lines available in the public transit dataset. \it{Note: The blue marker indicates NTNU's Gløshaugen campus. Due to the wide geographical spread, some outlying stops are not visible within the map bounds.}}
    \label{fig:home-stops-lines}
\end{figure}

\autoref{fig:home-stops-lines} presents a heatmap illustrating the approximate residential distribution of survey respondents, inferred from their self-reported nearest bus stops. For context, the top-rightmost stop, ``Dullumfeltet'', is located approximately 37.9 km by road from the Gløshaugen campus (indicated by the blue marker). At the other extreme, the bottom-leftmost stop, ``Fannrem stasjon'', lies around 46.9 km away.

\subsection{Quantitative: Big Data Analysis}
\label{sec:methodology-quantitative}
The qualitative survey was designed to gather mobility-related information on: \it{(i)} the nearest bus stops to respondents' homes and campus, \it{(ii)} commuting options used during the summer and winter months, and \it{(iii)} start and end times at work, both before and after the reduction in parking availability.
Said information is what provides the link between the qualitative and quantitative approaches used in this study, allowing us to study the spatiotemporal dynamics of human mobility flows in and around the main campus at Gløshaugen based on the stops, roads, times and modes indicated in the responses.

For the purposes of this study, the available sources of historical big mobility data include public transit data obtained from AtB and Entur; people flow data inferred from cellular signals; tollbooth data provided by the Norwegian Public Roads Administration (NPRA); and travel duration and route data collected through a variety of services and APIs.
There are two NPRA tollbooths of interest: one located north of ``Elgeseter Bru'' and the other along ``Elgeseter Gate'', near the ``Hesthagen'' bus stop. Said tollbooths can provide historical data on background vehicular flows in the vicinity of the campus.

\subsubsection{Public Transit Data}
\label{sec:public-transit-data}
This study uses data from AtB, Trondheim's public transit authority, collected through automated passenger counting (APC) systems installed on buses. These systems provide reliable records of passenger activity across the city's public bus transit network.
The dataset spans from 1st March 2022 to 30th November 2023 (post-COVID), covering 433,860 unique trips over 632 days, across 6 lines and 200 stops within Trondheim. As shown in the inset map in \autoref{fig:home-stops-lines}, the selected lines cover a broad section of the city, with all routes including stops near the Gløshaugen campus and a couple serving stops close to the Dragvoll campus.

The raw dataset was preprocessed to handle missing or erroneous passenger counts and timestamps, and to remove trips with incomplete stop sequences. The cleaned dataset includes 47 attributes, capturing stop-level arrival and departure times, stop locations and names, boarding and alighting counts, trip-level operational parameters, and additional engineered features such as terrain, land use and local demographics. This provides a comprehensive basis for analysing transit flows and accessibility in and around Gløshaugen.

\begin{figure}[tb!]
    \centering \includegraphics[width=.99\linewidth]{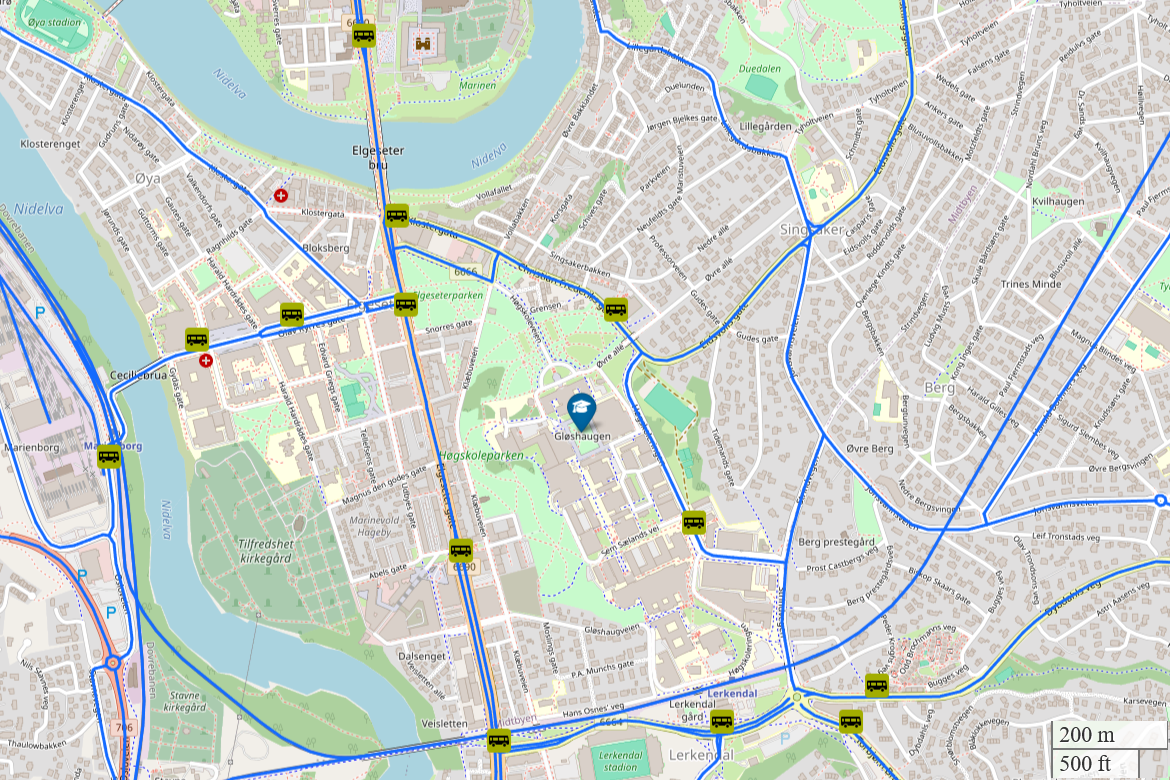}
    \caption{Overview of bus stops and major roads around Gløshaugen with available historical mobility data. \it{Note: The blue marker indicates NTNU's Gløshaugen campus. Road segments with available historical \it{peopleFlow} data are highlighted in blue.}}
    \label{fig:bd-data-sources}
\end{figure}

\subsubsection{Cellular Network Data}
\label{sec:cellular-network-data}
Network data derived from the activity of millions of mobile subscribers can provide valuable real-time and historical insights into the spatiotemporal dynamics of population flows within and across regions.
In this study, we use anonymised mobility data from Telia's ``Crowd Insights'' platform \cite{Telia2021cim}, specifically in the form of routing reports that model \it{peopleFlow}: aggregated travel behaviours inferred from probable routes taken by groups of people, revealing preferences in routes and modes over various regions and times.

The routing data also span from 1st March 2022 to 30th November 2023 (post-COVID), with observations at hourly intervals across 637 days. The raw dataset was preprocessed to address the gaps caused by privacy-preserving aggregation thresholds.
The cleaned dataset includes temporal information, spatial identifiers for road segments via OpenStreetMap \it{wayIDs} and corresponding \it{peopleFlow} counts, which reflect movement intensity along each segment. In total, the dataset contains 1,986 unique \it{wayID} values, representing the road network as a series of connected segments with known coordinates.

\autoref{fig:bd-data-sources} shows the bus stops and major roads used by the survey respondents when arriving and departing from the main campus at Gløshaugen. The road segments with available  historical \it{peopleFlow} data highlighted in blue, enabling the analysis of route preferences, modal shifts, and the spatial distribution of flows around Gløshaugen.

\subsubsection{Travel Durations and Routes}
\label{sec:travel-durations-routes}
For each respondent, the survey captured spatial data on the nearest bus stops to home and campus, as well as temporal data on typical start and end times at work. Together, these allowed individual analyses of travel duration, distance, and route requirements across various mobility modes, ensuring that respondents could arrive at campus at their specified start time (``morning'') and depart at their end time (``afternoon'').
To support this, we used the TravelTime platform \citep{TravelTime2025tad,Van_Rees2025htr} through the Python API (application programming interface) to perform commuting analyses for each respondent, considering the following modes:

\begin{itemize} \itemsep=0pt
    \item \it{Driving:} This encompasses driving routes, including private cars, carpooling and shared commutes, with ferry options where relevant. The API query was done using \mt{Ferry(type="driving+ferry")}, which allows for driving routes that include ferry crossings.
    \item \it{Transit:} This includes bus, ferry, and rail routes, with transfers allowed. The API query was performed using \mt{PublicTransport(max_changes=transfers)}, where \mt{transfers} is a parameter that limits the number of transfers to 5, according to survey responses.
    \item \it{Cycling:} This covers active mobility routes, including the use of ferries where applicable for home--campus commutes across the Trondheim fjord. The API query was done using \mt{Ferry(type="cycling+ferry")}, which allows for cycling routes that include ferry crossings.
\end{itemize}

\section{Results and Discussion}
\label{sec:results-discussion}
This section reports the main findings of the qualitative survey and the quantitative analysis of mobility data. The results are presented in two parts: first, the experiences and challenges reported by NTNU employees and students in response to parking reduction policies in \autoref{sec:results-qualitative}; second, the broader commuting patterns and trends identified through large-scale mobility datasets in \autoref{sec:results-quantitative}; finally, the qualitative and quantitative findings are integrated in \autoref{sec:results-mobility} to provide a comprehensive overview of commuting dynamics at Gløshaugen. Together, these findings inform the discussion of the impacts and implications of campus mobility reforms.

\subsection{Qualitative Survey}
\label{sec:results-qualitative}
This section presents a synthesis of the qualitative findings of the quantitative survey, highlighting the perspectives and experiences of students and employees in response to campus parking reductions. Drawing on open-ended survey responses, the analysis explores the main themes and concerns raised by the participants, providing context and depth to the quantitative results discussed in \autoref{sec:results-quantitative} and the mobility patterns analysed in \autoref{sec:results-mobility}.

\subsubsection{Quantitative Context and Key Findings}
\label{sec:results-context-findings}
\begin{figure}[tb!]
    \centering
    \includegraphics[width=.99\textwidth]{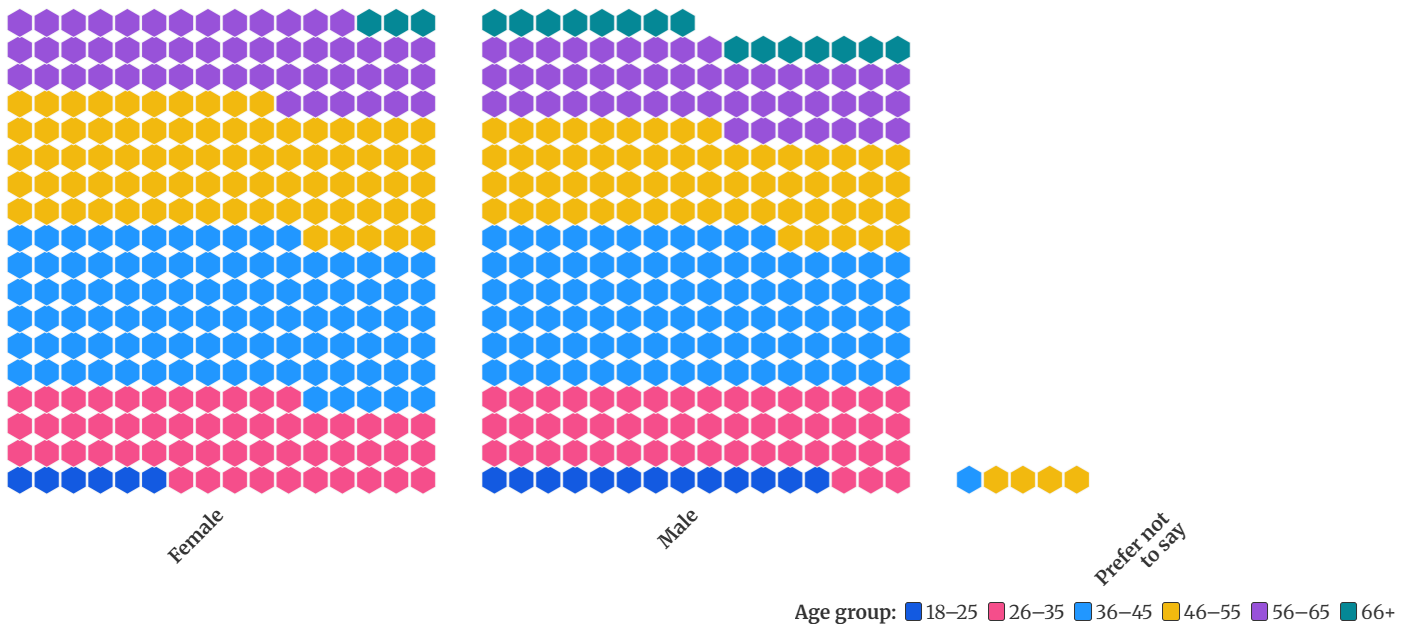}
    \caption{Demographic breakdown of the 573 survey respondents by gender and age group, highlighting the nearly balanced gender distribution and the concentration of respondents in the 36--55 age range.}
    \label{fig:demographics-age-gender}
\end{figure}

The qualitative survey received 573 responses, providing a robust dataset to understand commuting patterns, demographics, and the anticipated impact of parking reduction policies. The sample was nearly gender-balanced (50.3\% female, 48.9\% male) as shown in \autoref{fig:demographics-age-gender} and included a wide range of roles, with administrative employees (47.5\%) and permanent faculty (28.6\%) being the largest groups. The age distribution was concentrated in the 36--55 range (58.1\%), and a majority (59.3\%) had children, underlining the importance of family logistics in commuting decisions.

\begin{figure}[tb!]
    \centering
    \begin{subfigure}[b]{.99\textwidth}
        \centering \includegraphics[width=\textwidth]{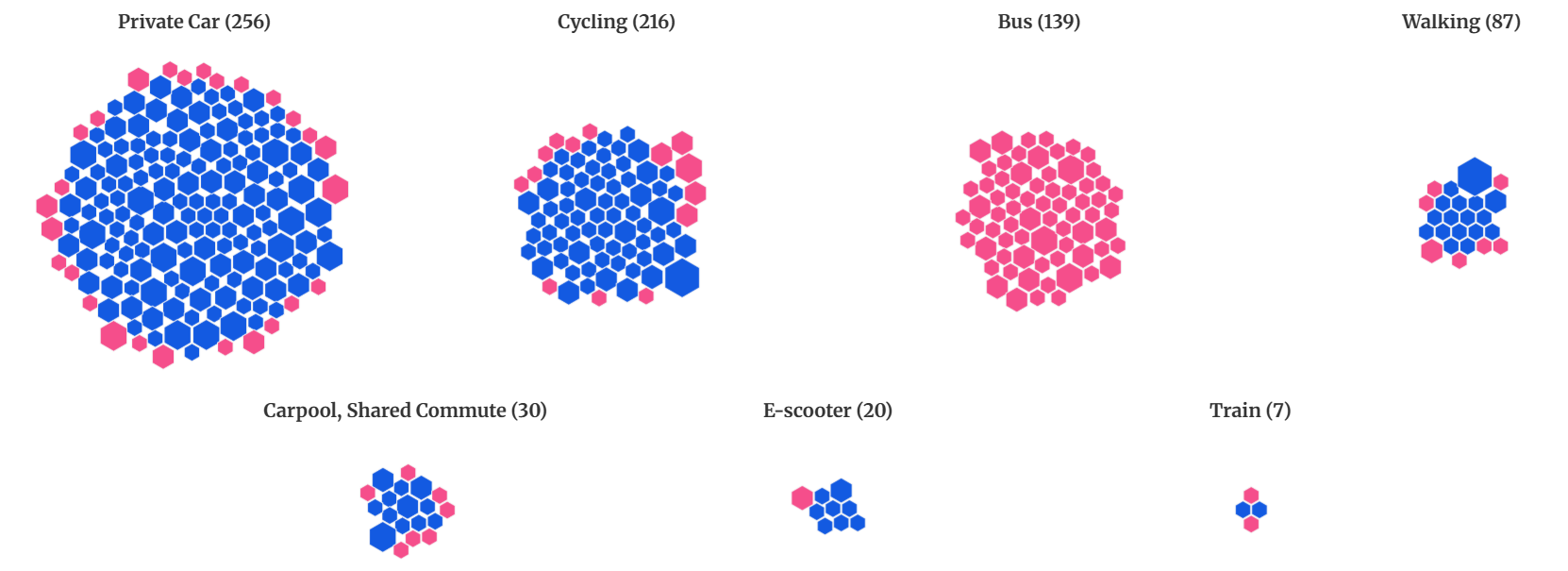}
        \caption{Summer} \label{fig:commute-modes-transfers-summer}
    \end{subfigure}
    \begin{subfigure}[b]{.99\textwidth}
        \centering \includegraphics[width=\textwidth]{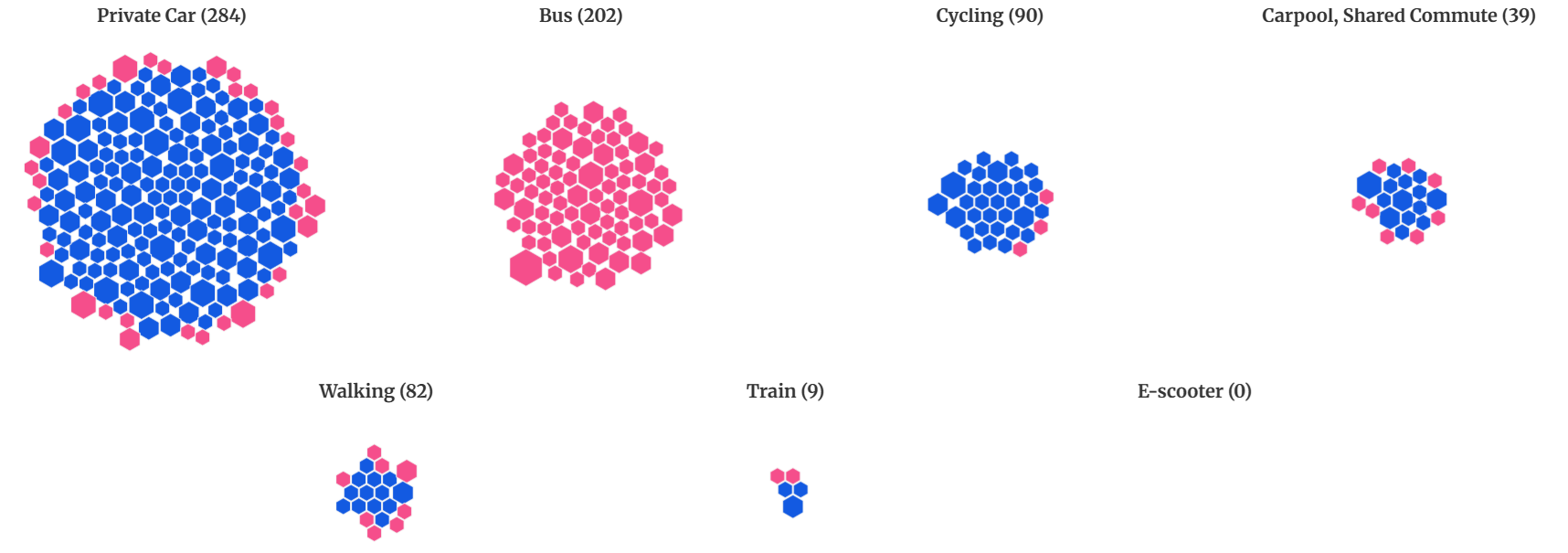}
        \caption{Winter} \label{fig:commute-modes-transfers-winter}
    \end{subfigure}
    \caption{Overview of mobility modes used by survey respondents during summer and winter, with bus users highlighted in red. \it{Note: Point size indicates the number of bus transfers per trip (0--5); larger points represent more transfers.}}
    \label{fig:commute-modes-transfers}
\end{figure}

Campus attendance was high, with 72.6\% of the respondents present on campus five or more days a week, and only 1.9\% attending one or two days. The use of private cars was the dominant commute mode, especially in winter (49.4\%), but cycling (37.2\% in summer) and bus (34.9\% in winter) were also significant, as shown in \autoref{fig:commute-modes-transfers-summer} and \autoref{fig:commute-modes-transfers-winter}. Walking, carpooling, and e-scooters played a smaller role, with e-scooter use dropping to zero in winter.

The average number of bus transfers per trip was 0.72, indicating that many commutes involved at least one transfer.
While 73.3\% saw the bus as a feasible alternative, only 37.2\% considered cycling and 21.1\% walking, reflecting both interest and perceived barriers. In particular, 42.1\% anticipated increased mobility costs, and 29.8\% found the current bicycle and active travel facilities insufficient (with 27.9\% neutral).

\begin{figure}[tb!]
    \centering
    \includegraphics[width=.99\textwidth]{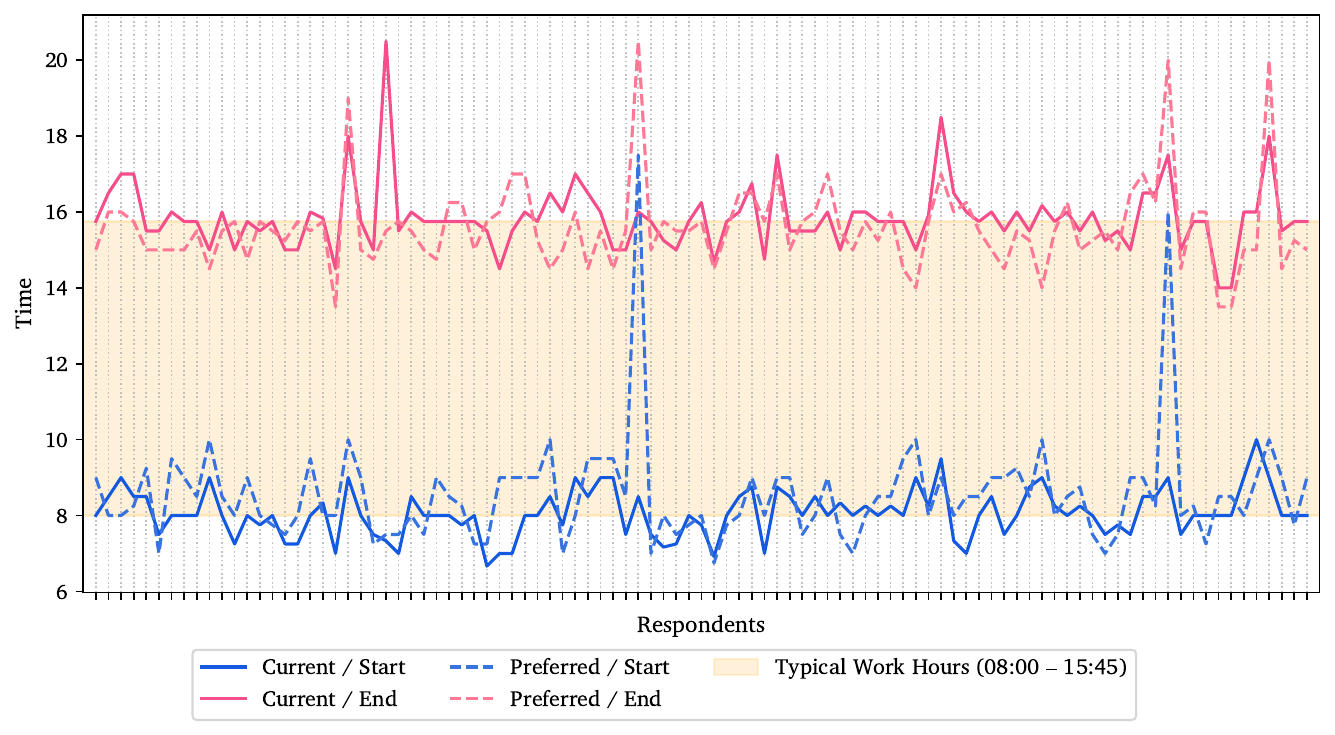}
    \caption{Potential changes to work start and end times among 97 respondents who anticipate adjusting their work patterns due to parking reduction. \it{Note: Solid lines indicate current work hours; dashed lines indicate anticipated changes; and the shaded area represents the typical workday (08:00--15:45).}}
    \label{fig:work-hours-change-combined}
\end{figure}

\begin{figure}[tb!]
    \centering
    \includegraphics[width=.99\textwidth]{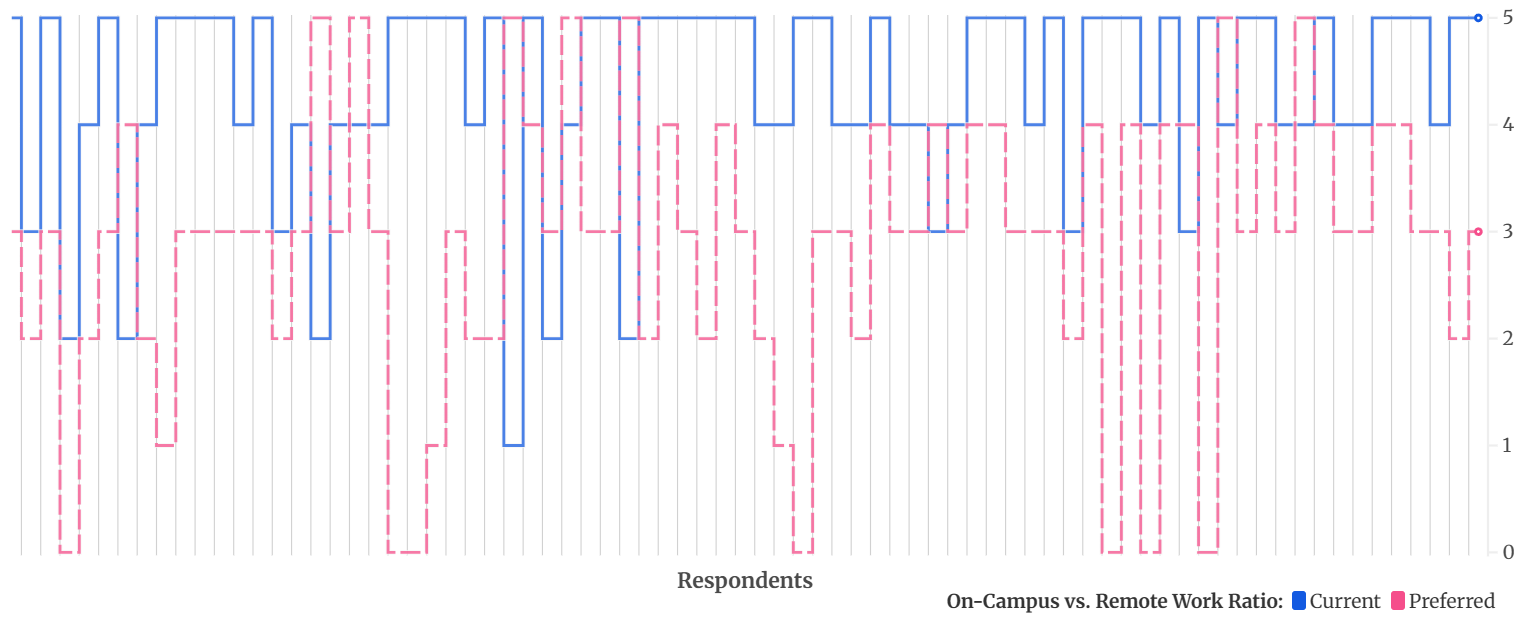}
    \caption{Potential changes in the on-campus to remote work ratio among 77 respondents who anticipate adjusting their work patterns due to parking reduction. \it{Note: A lower ratio indicates reduced on-campus presence and a shift toward more remote work.}}
    \label{fig:campus-remote-ratio}
\end{figure}

Family and daily schedule factors were prominent: 59.3\% of the respondents had children, with an average workday that started at 08:00 and ended at 15:45. The average on-campus to remote work ratio was 4.58 (on a 0--5 scale, where 5 means fully on-campus), indicating a strong presence on-site.
\autoref{fig:work-hours-change-combined} and \autoref{fig:campus-remote-ratio} illustrate the significant impact of parking reduction on work schedules, with many employees expecting to change their work hours and reduce their presence on campus to accommodate longer commutes or reduced parking availability.

Of the 286 respondents who indicated possible or definite changes to their work schedule, 97 also reported potential adjustments to their start and end times. \autoref{fig:work-hours-change-combined} presents the distribution of these individuals, with 64.9\% and 67.0\% anticipating later starts and earlier finishes, respectively.
Thus, the average workday across the said 97 respondents would shift from 08:03--15:50 (7h 47m) to 08:34--15:36 (7h 02m), effectively shortening the workday by 45 minutes and reducing overlap with peak commuting hours.

Similarly, \autoref{fig:campus-remote-ratio} shows potential adjustments to the on-campus to remote work ratio indicated by 77 (13.4\%) of the 573 respondents. Of these, 66 (85.7\%) expected to decrease their presence on campus, thus decreasing the average on-campus to remote work ratio from 4.58 to 4.39 among all respondents.
When the 77 respondents who anticipated changes are considered alone, the average ratio drops from 4.31 to 2.88, indicating a substantial shift toward remote work.

\begin{figure}[tb!]
    \centering
    \includegraphics[width=.99\textwidth]{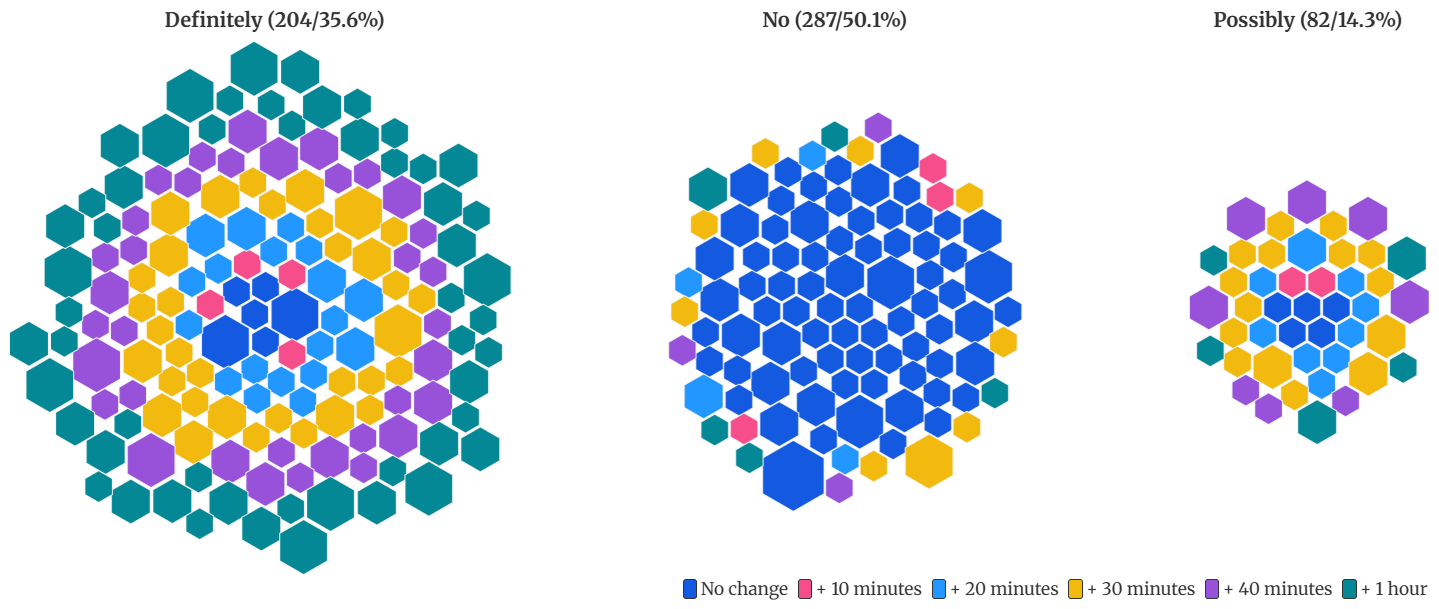}
    \caption{Distribution of respondents indicating whether their work schedule would be affected by parking reduction, with estimated additional commute time visualized. \it{Note: Point size indicates the number of bus transfers per trip (0--5); larger points represent more transfers.}}
    \label{fig:parking-commute-time-transfers}
\end{figure}

\autoref{fig:parking-commute-time-transfers} illustrates the distribution of respondents' expectations regarding how parking reduction would affect their work schedule and commute time. The majority (50.1\%) did not expect their work schedule to change, while 35.6\% anticipated definite impacts and 14.3\% were uncertain. Commute time expectations varied, with 45.2\% predicting no change, but 13.1\% expecting an additional hour or more due to parking reductions.

The potential for employee attrition was also evident: 15.7\% would consider leaving NTNU due to parking reductions, and 22.5\% were unsure. The most common difficulties reported were increased travel duration (86.0\%), additional costs (83.8\%), weather (82.9\%), schedule conflicts (82.2\%), and family logistics (81.2\%).

\begin{figure}[tb!]
    \includegraphics[width=.99\textwidth]{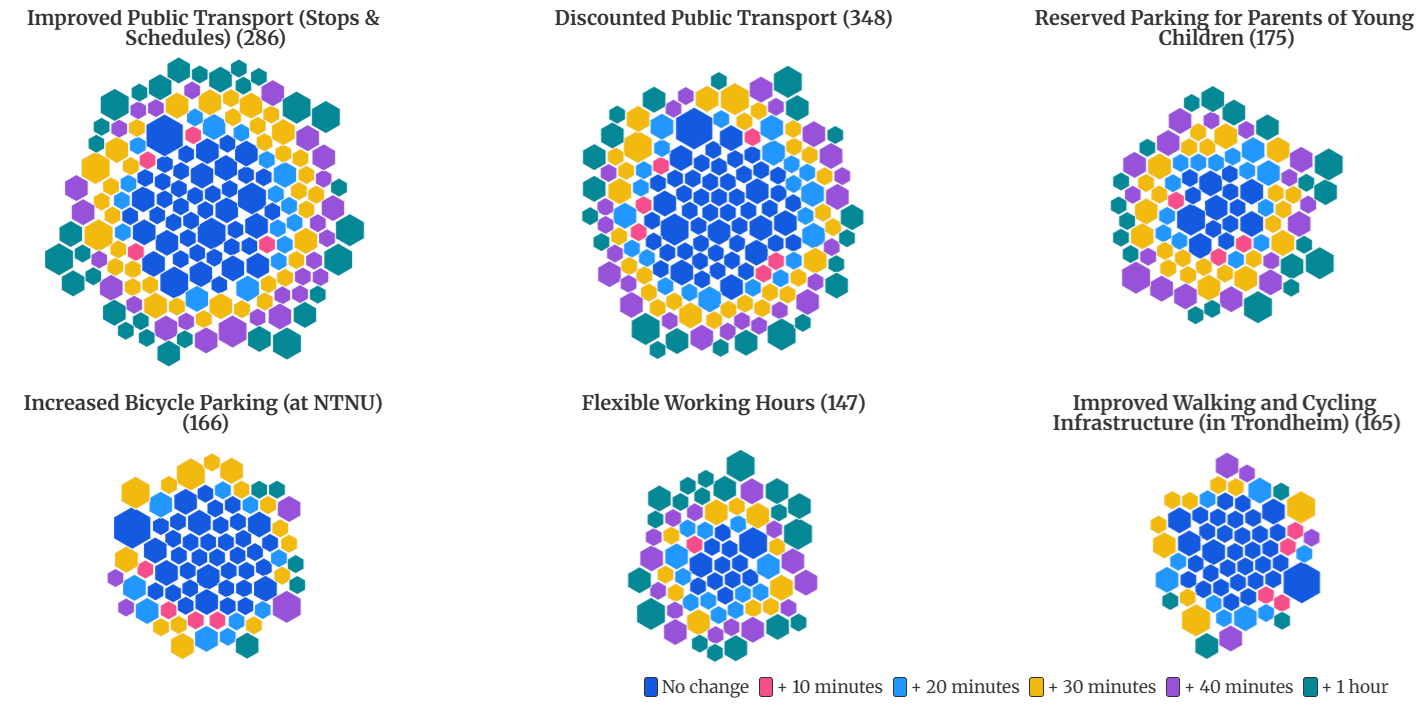}
    \caption{Preferred incentives to mitigate the impact of parking reduction on commuting, with estimated additional commute time visualized. \it{Note: Point size indicates the number of bus transfers per trip (0--5); larger points represent more transfers.}}
    \label{fig:incentives-commute-time-transfers}
\end{figure}

\autoref{fig:incentives-commute-time-transfers} reveals that the most preferred incentives to mitigate the impacts of parking reduction were: discounted public transit (60.7\%), improved stops and schedules (49.9\%), reserved parking for parents of young children (30.5\%), increased bicycle parking at NTNU (29.0\%), and improved walking and cycling infrastructure in Trondheim (28.8\%). Flexible working hours and other suggestions were also noted. These findings set the stage for understanding the qualitative feedback, which provides deeper insight into the lived experiences behind these statistics.

\subsubsection{Thematic Analysis of Open Text Responses}
\label{sec:results-thematic-analysis}
The open text responses in the survey were thematically analysed and grouped into three main categories: difficulties, incentives, and comments. These categories reflect the lived experiences, practical challenges, and emotional responses of NTNU employees and students to the parking reduction policy. The following synthesis draws on detailed participant feedback to illustrate the complexity and diversity of perspectives.

\paragraph{Difficulties}
A substantial number of respondents described significant and often intersecting obstacles resulting from the parking reduction. These challenges affected daily routines, work productivity, and general well-being, and were especially acute for those with family or health-related responsibilities. Key difficulties included:

\begin{itemize}
    \item \textbf{Increased Commuting Time and Productivity Loss}: Many participants reported that switching from the car to public transit doubled or even tripled their daily commute, some facing more than three hours of travel each day. Night-shift workers and those with multiple transfers were particularly affected, and some considered changing jobs due to unmanageable commuting demands. Lost time also led to more remote work, reduced collaboration, and the need to work on evenings or weekends to compensate.
    \item \textbf{Family and Childcare Constraints}: Parents, especially those with young children or dependents with special needs, found public transit incompatible with their schedules. Early dropoffs, late kindergarten openings, and the need for flexible on-demand commuting were cited as reasons why car access remains essential. Some described extreme trade-offs, such as dropping children off before kindergartens open to give early lectures.
    \item \textbf{Health and Mobility Barriers}: Respondents with chronic illnesses, disabilities, or temporary injuries noted that public transit and cycling are not viable options for them. Some described being denied accessible parking despite having medical needs, and others noted that standing on crowded buses or cycling in winter was physically challenging. Wheelchair users and those with partial disabilities reported feeling overlooked by the current system.
    \item \textbf{Public Transit Limitations}: Many described the bus system as unreliable, overcrowded, and inflexible, especially in winter or for those who live far from the stops. Indirect routes, infrequent service, and long transfer times made public transit impractical for a significant portion of the community. Some noted the lack of service on weekends and the ``milk run'' nature of local buses, which made even short trips much longer than by car.
    \item \textbf{Cycling Infrastructure Gaps}: While there was enthusiasm for cycling, respondents highlighted the lack of secure indoor bike parking, poor winter maintenance of bike paths, and insufficient facilities for cleaning and drying bikes. Some stopped biking altogether in winter due to unsafe conditions. The requests included drying rooms, indoor parking near the offices, and accessible showers.
    \item \textbf{Reduced Flexibility and Quality of Life}: The loss of parking was seen as reducing daily flexibility, making it harder to run errands, attend appointments, or transport equipment to and from campus for field work. Some anticipated early retirement or job changes as a result. Specific work-related difficulties included moving tools, visiting partner sites, or shuttling between campuses, all of which became more challenging without a car.
\end{itemize}

\begin{figure}[tb!]
    \centering
    \includegraphics[width=.99\textwidth]{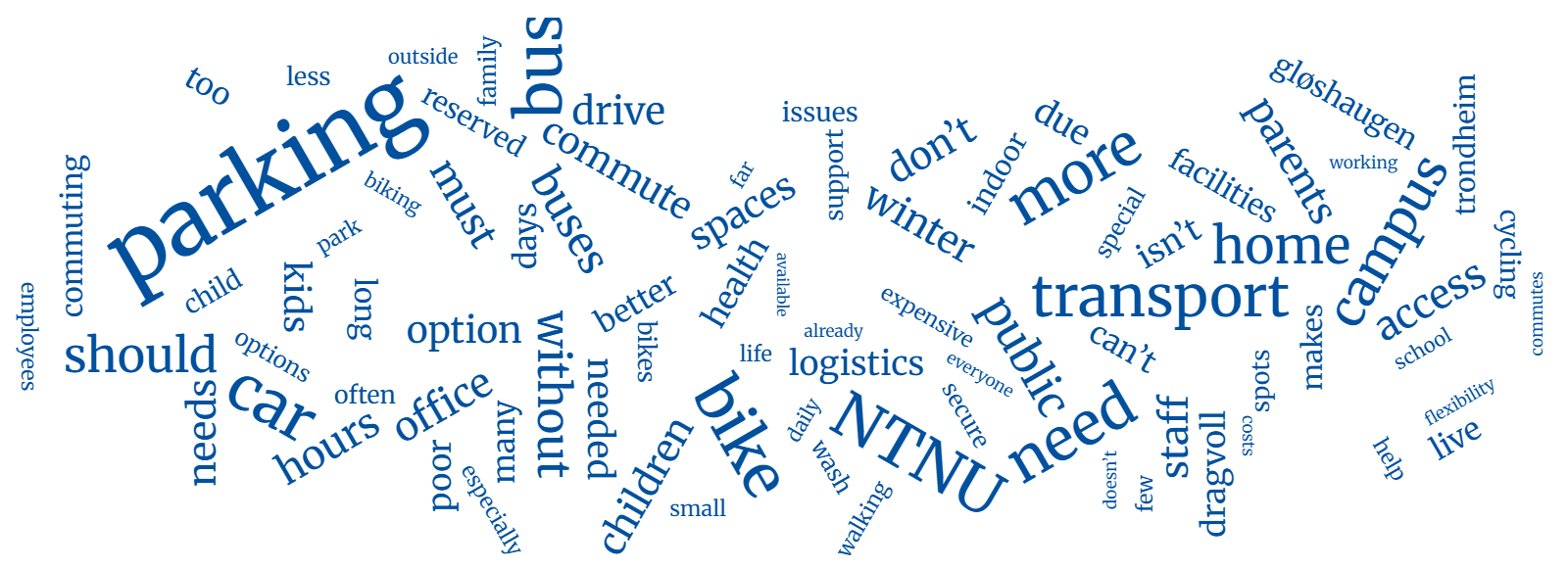}
    \caption{Word cloud of the most frequently used terms in open text responses, highlighting key themes and concerns.}
    \label{fig:free-text-wordcloud}
\end{figure}

\paragraph{Incentives}
The respondents offered a variety of suggestions to mitigate the negative impacts of parking reduction and to support more sustainable commuting. These incentives reflect both practical needs and a desire for more inclusive policy-making:

\begin{itemize}
    \item \textbf{Prioritised Parking Access}: Many called for restoring or increasing parking for employees with long commutes, health conditions, or caregiving responsibilities. Suggestions included distance-based prioritisation, reserved employee parking, and flexible options like occasional-use day passes or clip cards. Some suggested that it would be fair for employees to have different access compared to students.
    \item \textbf{Remote Work Opportunities}: There was strong support for expanding and formalising remote work policies, especially for those with family obligations, health limitations, or long commutes. Respondents noted that current implementation is inconsistent between departments, and some suggested rotating in-office days or assigning home office quotas.
    \item \textbf{Improved Cycling Facilities}: The requests included secure indoor bike storage (including locked cages and climate-controlled facilities), maintenance and cleaning facilities, and better access to showers and lockers. E-bike subsidies or leasing programmes and winter maintenance of bike paths were also frequently mentioned.
    \item \textbf{Enhanced Public Transit and Shuttle Services}: Suggestions included more direct and reliable bus routes (avoiding the city centre), shuttle services between campuses and external hubs, and more park-and-ride facilities. Some advocated for counting commute time as work time for technical employees who cannot work remotely.
    \item \textbf{Additional Support Measures}: The proposals included increased salary or stipends to offset increased commuting costs, better winter maintenance of sidewalks and paths, guest parking, and more inclusive, transparent policy-making that considers diverse life situations. Flexible parking options for irregular needs (e.g. medical appointments, heavy equipment transport) were also requested.
\end{itemize}

\paragraph{Comments}
The comments section captured a wide range of emotional reactions, critiques, and further suggestions. Many respondents expressed frustration and concern about the perceived lack of inclusivity and transparency in the policy process. Key themes, as illustrated by the word cloud in \autoref{fig:free-text-wordcloud}, included:

\begin{itemize}
    \item \textbf{Frustration and Attrition Risk}: Many described the parking reduction as a significant challenge, leading to thoughts of remote work, early retirement, or leaving NTNU. There was a strong sense that the needs of the employees and their voices were not being heard. For some, parking was the last remaining perk and its removal symbolised a broader erosion of trust and morale.
    \item \textbf{Complex Family and Care Responsibilities}: The respondents emphasised that car access is essential not only for parents of small children but also for those with broader caregiving duties, such as supporting elderly or disabled relatives. A policy focused only on ``small children'' would still exclude many in need.
    \item \textbf{Public Transit Not a Universal Solution}: Detailed accounts explained why buses and trains are not realistic alternatives for many, due to long delays, poor or cancelled service, and overcrowding, resulting in less time at home and reduced work performance. This led to a strong shift towards home office or attrition.
    \item \textbf{Health and Accessibility Concerns}: Calls for guaranteed parking for people with chronic or temporary health challenges were common, with criticism that current policies do not adequately address a range of mobility needs. Some called for prioritisation that accounts for more than just physical disability badges, including episodic illness or physical strain from commuting with heavy loads.
    \item \textbf{Cycling Infrastructure Needs}: Cyclists expressed a positive outlook, but stressed the need for secure parking, washing stations, and changing facilities to make cycling feasible throughout the year. Some noted that NTNU’s current facilities are far behind other Nordic universities and could be upgraded with basic investments.
    \item \textbf{Policy Process and Equity Critique}: Many criticised the policy as ideologically driven and insufficiently based on data or lived experience. There were calls for more transparent, inclusive, and socially sustainable decision-making at NTNU. Some respondents claimed that the survey was biased or methodologically flawed and that real accessibility needs, especially during difficult life phases, are being ignored.
\end{itemize}

\subsubsection{Summary and Implications}
The responses to the qualitative survey reveal a complex landscape of commuting challenges and opportunities for improvement. Although there is a broad support for sustainable mobility goals, respondents expressed concerns about the pace and design of parking reduction policies and their potential impacts on employee well-being, equity, and retention. The difficulties expressed highlight the considerations for a more inclusive approach that accounts for diverse commuting needs, particularly for those with family responsibilities, health conditions, or mobility challenges.

The quantitative findings underscore these concerns: with 59.3\% of the respondents having children and 72.6\% attending campus five or more days per week, family logistics and regular campus presence create significant mobility constraints. The dominance of private car use, especially in winter (49.4\%), reflects reported practical necessities rather than a mere preference. The fact that 42.1\% anticipate increased mobility costs and 15.7\% would consider leaving NTNU demonstrates the serious financial and career implications of these policy changes. Most strikingly, 86.0\% identify increased travel duration as a primary difficulty, highlighting how parking reduction may affect work-life balance for a substantial portion of the campus community.

\vspace{\baselineskip}

\subsection{Quantitative Analysis}
\label{sec:results-quantitative}
This section presents the quantitative analysis of mobility patterns around Gløshaugen, complementing the survey findings with objective data on actual travel behaviours. As described in \autoref{sec:methodology-quantitative}, we analyse big mobility data from two complementary sources: cellular network-based routing data and public transit passenger counts.
These data sources help us understand the current mobility patterns, complementing the subsequent mobility analysis in \autoref{sec:results-mobility}, and how they may change due to increased demand for public transit services and reduced parking availability after the campus consolidation project.

\subsubsection{Public Transit Usage at Campus Stops}
\label{sec:results-public-transit}
With 73.3\% of survey respondents identifying public transit as a feasible alternative to driving in \autoref{sec:results-context-findings}, understanding current transit usage patterns is crucial for contextualising the mobility landscape in NTNU.
Using respondents' self-reported nearest bus stops to campus and filtering for those used by three or more respondents, we identified 20 bus stops that are relevant for this analysis.

Further filtering for stops for which we have access to historical public transit data (see \autoref{sec:public-transit-data}), we narrowed it down to 16 stops around the Gløshaugen and Dragvoll campuses.
These stops include major transit hubs such as ``Kongens gate'', ``Prinsens gate'', and ``Nidarosdommen'', which serve as key access points for students and employees commuting to campus.

To estimate the total volume of public transit usage at these stops, we calculated the sum of boardings and alightings for each stop across all trips in the dataset, resulting in the stop volumes, \it{stopVolume}. The \it{stopVolume} was subsequently aggregated by hour to reveal the overall usage patterns and identify the busiest stops and peak travel times.
The analysis focused on post-pandemic data from March 2022 to November 2023 and was restricted to workdays to capture typical commuting behaviour.

\begin{figure}[tb!]
    \centering
    \includegraphics[width=.99\textwidth]{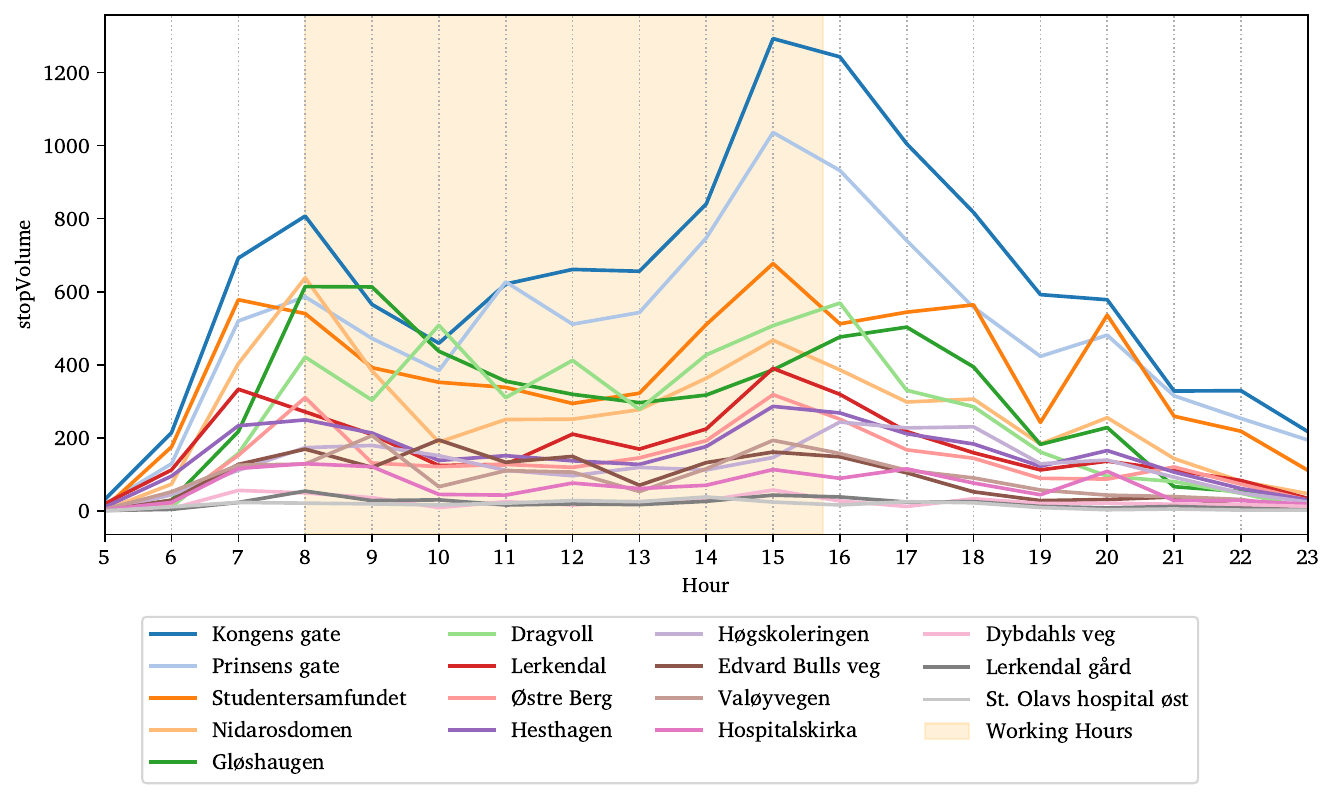}
    \caption{Total hourly boardings and alightings at all public transit stops near the Gløshaugen and Dragvoll NTNU campuses. Major stops such as ``Kongens gate'', ``Prinsens gate'', and Nidarosdommen dominate the morning peak, while usage is more dispersed across stops during other periods. \it{Note: The data is aggregated from the busiest day recorded, Wednesday, November 15, 2023.}}
    \label{fig:stopvolume-allstops}
\end{figure}

Our analysis revealed that the highest volume of public transit usage typically occurred in November, on Wednesdays around the 15:00 end-of-work rush hour.
Thus, the busiest day recorded was Wednesday, November 15, 2023, with a total of 61,180 boardings and alightings across all stops.
The hourly \it{stopVolume} aggregations data from that day, shown in \autoref{fig:stopvolume-allstops}, with the morning peak (08:00, 5,159 passengers) more spread out between the stops, while the afternoon peak was higher, more spread out between 15:00 (6,097 passengers) and 16:00 (5,674 passengers), and dominated by the stops at ``Kongens gate'' and ``Prinsens gate''.

The stops closest to the campuses, such as ``Gløshaugen'' and ``Dragvoll'', had their morning and afternoon peaks shifted by an hour, with the morning peak occurring between 08:00 and 09:00 and the afternoon peak between 16:00 and 17:00, or as late as 18:00 for ``Gløshaugen''.
While some of the survey respondents indicated they will be arriving later in the morning and leaving earlier in the afternoon, such potential changes had little impact on the average start and end times at work, which now occur around 08:00 and 16:00, respectively.

The analysis suggests that the challenges students and employees currently experience with public transit may persist with the consolidation of campuses and the reduction of parking facilities, other things being equal.
Using data from the busiest available day provides a snapshot of selected bus stops and routes operating at full capacity, providing a foundation for projecting the expected increase in demand from the additional 9,300 daily campus users \citep{Ramboll2024kfv,NTNU2025ocn}.
These projections should account for the 24.1\% and 34.9\% who currently use public transit in the summer and winter months, respectively, while excluding the 22.5\% and 15.7\% who may possibly or definitely leave NTNU due to increased commuting difficulties.

\subsubsection{Crowd Movement Patterns aroung Gløshaugen}
The analysis of crowd movement patterns and the background traffic around the main campus at Gløshaugen provides insights into how the anticipated increase in campus users may impact mobility flows.
When combined with vehicular traffic data from the NPRA tollbooths, this analysis can help identify potential bottlenecks, congestion, and safety issues as increasing numbers of vehicles, cyclists, and pedestrians converge on the campus.

We manually identified 14 key roads around Gløshaugen that will serve as primary access routes for students and employees, limiting the analysis to the roads for which we have access to historical cellular network data (see \autoref{sec:cellular-network-data} and \autoref{fig:bd-data-sources}) on crowd movement patterns, \it{peopleFlow}.
Some of these roads are major arteries with considerable lengths, such as ``Elgeseter Gate'' and ``Strindvegen'', which were split into segments to account for inflows and outflows at intersections to and from smaller adjoining roads.

Similar to the public transit analysis, post-COVID data from March 2022 to November 2023 were analysed with a focus on weekdays to capture typical commuting patterns. The \it{peopleFlow} count was aggregated by hour to reveal the overall usage patterns and identify the busiest roads and peak travel times.
This temporal analysis is particularly relevant given that the overwhelming number of survey respondents (72.6\%) report being present on campus five or more days a week.

The analysis revealed that the highest volume of \it{peopleFlow} typically occurred in September, on Fridays around the 15:00 end-of-work rush hour. This is understandable, as September marks the start of the academic year, when many students and employees return to campus after the summer break.
However, the busiest day recorded for the \it{peopleFlow} was Tuesday, May 17, 2022, which was a Constitution Day celebration holiday in Norway. The mobility patterns on this day were atypical, so we focused our analysis on the busiest day recorded in the public transit dataset, Wednesday, November 15, 2023.

\begin{figure}[tb!]
    \centering \includegraphics[width=.99\textwidth]{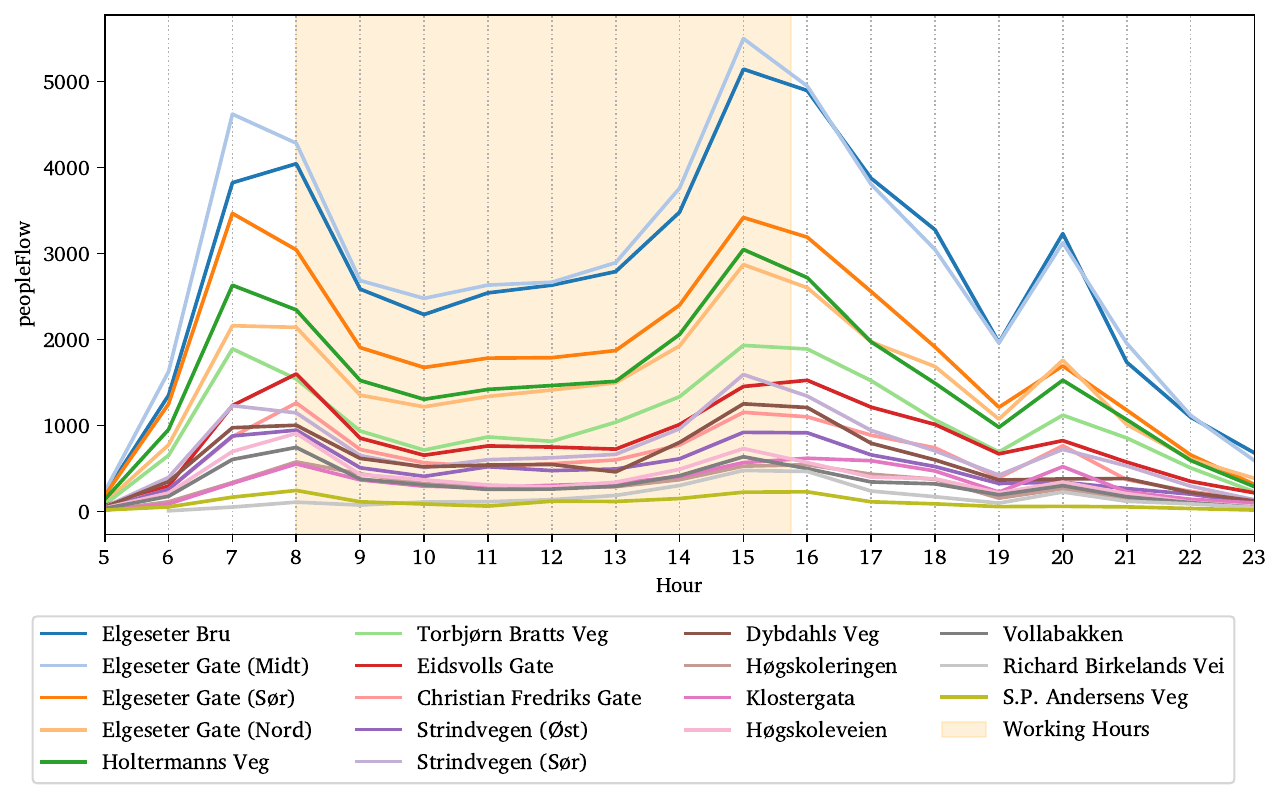}
    \caption{Total hourly crowd movement along selected roads surrounding Gløshaugen. The highest flows are observed on the main access routes (``Holtermanns Veg'', ``Elgeseter Gate'', and ``Elgeseter Bru'') with pronounced peaks during typical commuting hours. \it{Note: The data is aggregated from Wednesday, 15 November 2023, as in \autoref{fig:stopvolume-allstops}.}}
    \label{fig:peopleflow-allroads}
    \vspace{1em}
\end{figure}

On that day, the total \it{peopleFlow} across all selected roads was 309,448, with the highest flows occurring at ``Holtermanns Veg'', ``Elgeseter Gate'', and ``Elgeseter Bru'', as shown in \autoref{fig:peopleflow-allroads}. The hourly \it{peopleFlow} aggregations show trends similar to that of public transit usage at campus stops (see \autoref{fig:stopvolume-allstops}).
However, the morning peak was more spread out 07:00 (25,950 individuals) and 08:00 (26,478 individuals), while the afternoon peak was similarly more concentrated around 15:00 (31,438 individuals) and 16:00 (29,248 individuals). In addition, there was a slight increase in \it{peopleFlow} around 20:00.

These trends further confirm the temporal patterns observed with public transit and complement the analysis in \autoref{sec:results-public-transit} by providing a broader view of crowd movement patterns around Gløshaugen and the potential impact of the campus consolidation project and reduced parking availability on these mobility flows.
While the \it{peopleFlow} data does not distinguish between different mobility modes, it provides a useful proxy for understanding the overall movement patterns and potential congestion points around the campus.

\subsubsection{Summary and Implications}
These quantitative findings from the big mobility data, alongside those from the qualitative analysis conducted in \autoref{sec:results-qualitative}, highlight both the opportunities and challenges posed by the expected increase in campus users and the decline in parking availability.
To support more students and employees in transitioning to public transit and active mobility modes (within the limitations of their daily routines and circumstances), the existing mobility infrastructure and services around Gløshaugen would need to be adjusted to accommodate the increased demand.

The pronounced morning and afternoon peaks in public transit usage and crowd movement around Gløshaugen indicate potential congestion risks that could affect campus accessibility. Studies of bike-sharing systems in 15 European countries \citep{Waldner2025ddi} suggest similar temporal patterns for cycling: peak activity during commute hours with sustained afternoon use.
To mitigate these challenges, targeted interventions could be considered to improve public transit services, such as increasing bus frequency during peak hours, improving transfer coordination, and expanding cycling infrastructure.

\vspace{\baselineskip}

\subsection{Mobility Analysis}
\label{sec:results-mobility}
This section analyses commuting patterns between home and campus locations, combining the qualitative survey responses from \autoref{sec:results-qualitative} with available quantitative mobility data from \autoref{sec:results-quantitative}, supported by visualisations that highlight key behavioural trends.
The analysis examines home--campus commuting patterns, including travel flows between stop pairs, usage of different mobility modes, commuting distances and travel durations, and a congestion analysis comparing driving and cycling.

\subsubsection{Directed Flows between Home and Campus Locations}
\label{sec:results-directed-pairs}
The following analysis examines the directed flows between home and campus stops, with the aim of providing insight into the commuting routes and travel distribution across different locations.
The data is derived from responses to the \it{``Nearest bus stop to home''} and \it{``Nearest bus stop to campus''} questions in the ``Commute and Transport'' section of the qualitative survey (see \autoref{sec:methodology-qualitative}).

\begin{figure}[tb!]
    \centering
    \includegraphics[width=.99\textwidth]{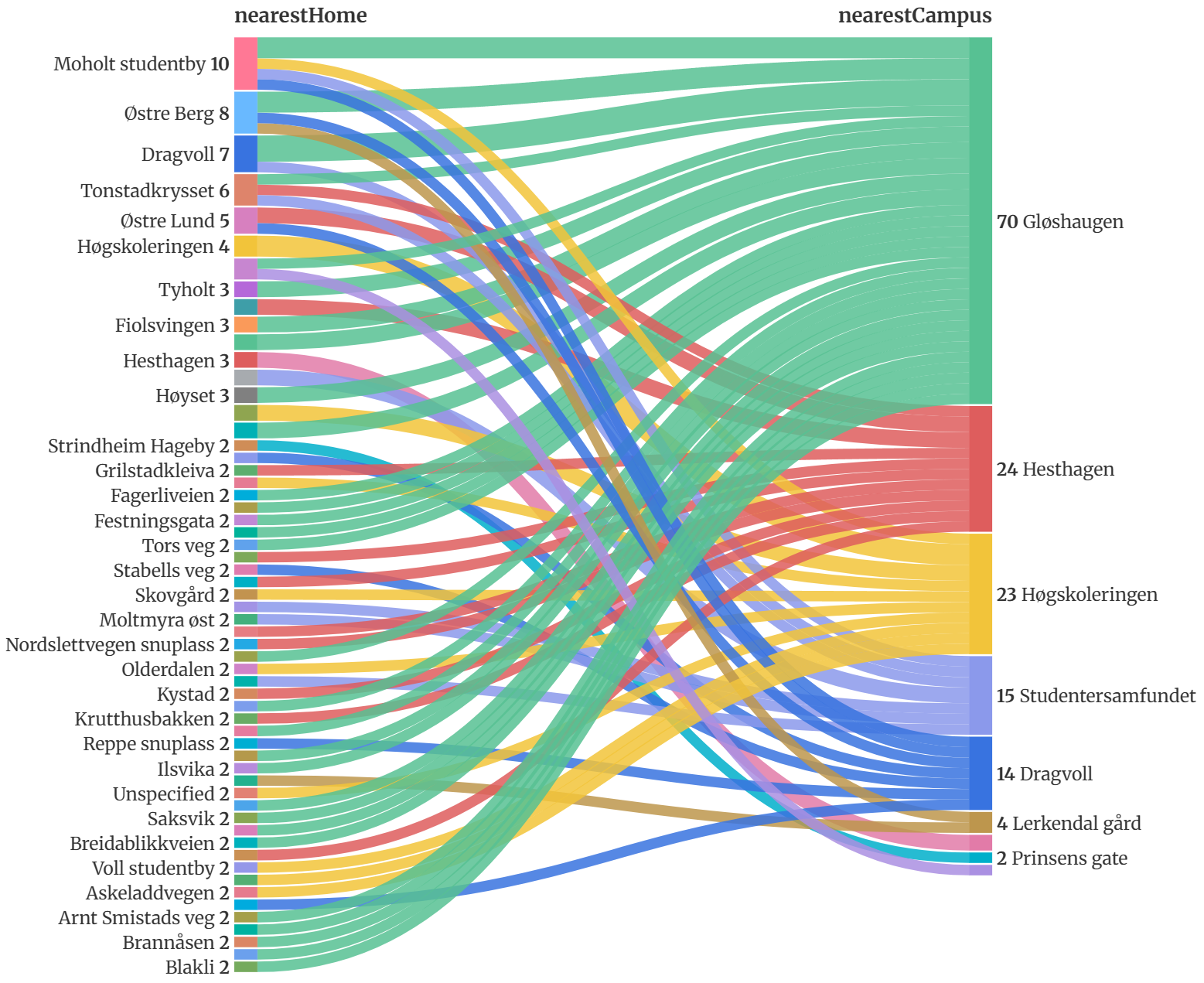}
    \caption{Sankey diagram showing directed flows between home and campus stops based on survey responses. The width of the bands corresponds to the number of trips between specific stop pairs. \it{Note: The visualisation is limited to high-volume home--campus stop pairs with more than one trip (covering 27.6\% of all respondents) for clarity.}}
    \label{fig:survey-stops-directed-pairs}
\end{figure}

The nearest bus stop to home serves as a proxy for the respondent's residential location, enabling the analysis of commuting patterns at the individual level while minimising the risk of divulging personally identifiable information (PII) in the otherwise anonymous survey.
The Sankey diagram in \autoref{fig:survey-stops-directed-pairs} shows the most common home--campus commuting corridors among survey respondents. Each band in the diagram represents a group of respondents, with the width of the bands corresponding to the number of trips between specific home--campus stop pairs.

The visualisation is limited to home--campus stop pairs with more than one trip for clarity, resulting in 59 unique home stops and 9 unique campus stops being displayed. These 59 home stops are distributed across the city of Trondheim and beyond, representing 157 (27.6\%) of the total of 573 survey respondents (and their commutes). However, the 9 campus stops in the diagram are used by 473 (82.5\%) of all respondents for their commutes.
The self-connections at the ``Gløshaugen'' and ``Høgskoleringen'' stops indicate respondents who live near the main campus at Gløshaugen, where the home and campus stops are the same.

With only the most significant flows included, the visualisation clearly shows that the campus stops most frequently used are those serving the Gløshaugen campus, such as ``Gløshaugen'', ``Hesthagen'', and ``Høgskoleringen''. The ``Dragvoll'' stop for the Dragvoll campus is also used by a notable number of respondents.
\autoref{fig:survey-stops-directed-pairs} offers a more detailed view of trip origins and destinations, serving as a useful complement to the overall stop volumes shown in \autoref{fig:stopvolume-allstops}.

\subsubsection{Mobility Modes: Seasonal Variations and Feasible Alternatives}
\label{sec:results-mobility-modes}
This section explores the variety of mobility modes used by the respondents for their commutes, taking into account seasonal variations and the perceived feasibility of alternatives to private car use.
The data presented in \autoref{tab:mobility-modes} is based on survey responses to a variety of multiple choice questions about primary mobility modes, their use for summer and winter commutes, and an evaluation of other modes as feasible alternatives to private car use.
The modes are ranked by their total frequency of use in both seasons. Modes that are only considered feasible alternatives are placed in a separate section at the bottom of the table, ranked by their total frequency as well.

\begin{table}[width=.80\linewidth,pos=tb!]
    \renewcommand{\arraystretch}{1.15}
    \caption{Mobility modes, their frequencies for summer and winter commutes, and their potential as feasible alternatives to private car use. \it{Note: The primary mobility modes are ranked by their total frequency of use across both seasons, while the feasible alternatives are only ranked by their total frequency as alternatives.}}
    \label{tab:mobility-modes}
    \begin{tabular*}{\tblwidth}{@{\extracolsep\fill} l|cc|cc@{} }
        \toprule
        Mobility Modes & Summer Commute & Winter Commute & Feasible Alternative \\
        \midrule
        Private Car & 256 & 284 & -- \\ Bus & 139 & 202 & 426 \\
        Cycling & 216 & 90 & 217 \\ Walking & 87 & 82 & 121 \\
        Carpool, Shared Commute & 30 & 39 & 36 \\ E-scooter & 20 & -- & 47 \\
        Train & 7 & 9 & 17 \\ Express Boat & 4 & 4 & 2 \\
        Running & 2 & 2 & 4 \\ Motorcycle & 4 & -- & 3 \\
        Moped & 3 & 1 & -- \\ Ferry & 2 & 2 & -- \\ Skis & -- & 1 & -- \\
        \midrule
        \it{No Relevant Alternatives} & -- & -- & 38 \\ \it{Unspecified} & -- & -- & 11 \\
        Taxi & -- & -- & 4 \\ Scooter & -- & -- & 2 \\ Tram & -- & -- & 1 \\
        \bottomrule
    \end{tabular*}
\end{table}

The table shows that the most frequently used mobility mode among respondents is the private car (36.3\%), followed by public transit (22.9\%), and active modes such as cycling (20.6\%) and walking (11.4\%).
The data also reveal a significant seasonal variation in the choice of mobility modes, with cycling usage dropping by 58.3\% from summer to winter, while e-scooters are only used in summer due to service restrictions in winter.
In contrast, public transit and private car usage increased by 45.3\% and 10.9\%, respectively, during the winter months, indicating a shift towards more sheltered and reliable modes of transport in colder weather.

Although most of the respondents consider public transit and active mobility feasible alternatives to the use of private cars, their actual usage rates remain relatively low. For example, although 45.9\% of the respondents view public transit as a feasible alternative, only 22.9\% actually use it for their commutes.
In addition, the low response rates for carpooling and shared commutes, whether as primary modes or feasible alternatives, indicate that the motivation behind the use of private cars (and the demand for parking) may be related to the practical realities of daily mobility and individual spatiotemporal constraints, rather than solely cost savings or convenience considerations.

The survey respondents who primarily use private cars but consider public transit or active mobility as feasible alternatives, as well as those who indicate a lack of relevant alternatives, represent a potential target group for understanding the barriers to adopting more sustainable mobility modes and for designing interventions.
The thematic analysis in \autoref{sec:results-thematic-analysis} provides further insights into the specific challenges and barriers faced by such respondents, highlighting the need for targeted interventions that address their unique mobility needs and constraints.

\subsubsection{Mobility Modes: Distances and Travel Durations}
\label{sec:results-distances-durations}
This section analyses the commuting distances and travel durations associated with the mobility modes used by survey respondents, focusing on the routes between home and campus locations.
The information on the nearest bus stops to home and campus, combined with the working (start and end) times, allows us to construct each respondent's typical commuting routes. These routes include home--campus and campus--home trips, with distances and durations analysed for various mobility modes such as cycling, public transit, and driving (private car, carpooling, and shared commutes).

The routes are obtained from the TravelTime API, as described in \autoref{sec:travel-durations-routes}, for April 23, 2025, which was selected as a typical date for commuting analysis. The date was chosen to ensure that the analysis reflects typical commuting patterns, avoiding any anomalies due to holidays or special events.
The routing logic is such that the respondent arrives at the campus stop no later than their specified working start time and departs from the campus stop no earlier than their specified working end time.
In addition, in situations where the TravelTime query returns multiple results, the fastest route (i.e., the one with the shortest travel duration) is selected for analysis.

Of the 573 survey responses, several respondents were excluded from the subsequent routing analysis for various reasons. The reasons for exclusion are as follows:

\begin{itemize} \itemsep=0pt
    \item Some respondents (n=9) had incomplete or missing information on their home or campus stops, which prevents the routing analysis from being performed.
    \item For respondents who live in close proximity to their campus location and walk to/from work (n=17), their home and campus stops are the same, resulting in zero commuting distances. These cases are excluded from the routing analysis, as they do not provide meaningful data for distance or duration comparisons.
    \item There were some respondents for whom routing data for public transit could not be obtained (n=3). Due to poor transit connections, the routing API could not find a suitable public transit route for campus-bound or home-bound trips, making it impossible to compare the modes meaningfully. These cases were only excluded from comparative analyses between mobility modes.
\end{itemize}

\begin{figure}[tb!]
    \centering \includegraphics[width=.99\textwidth]{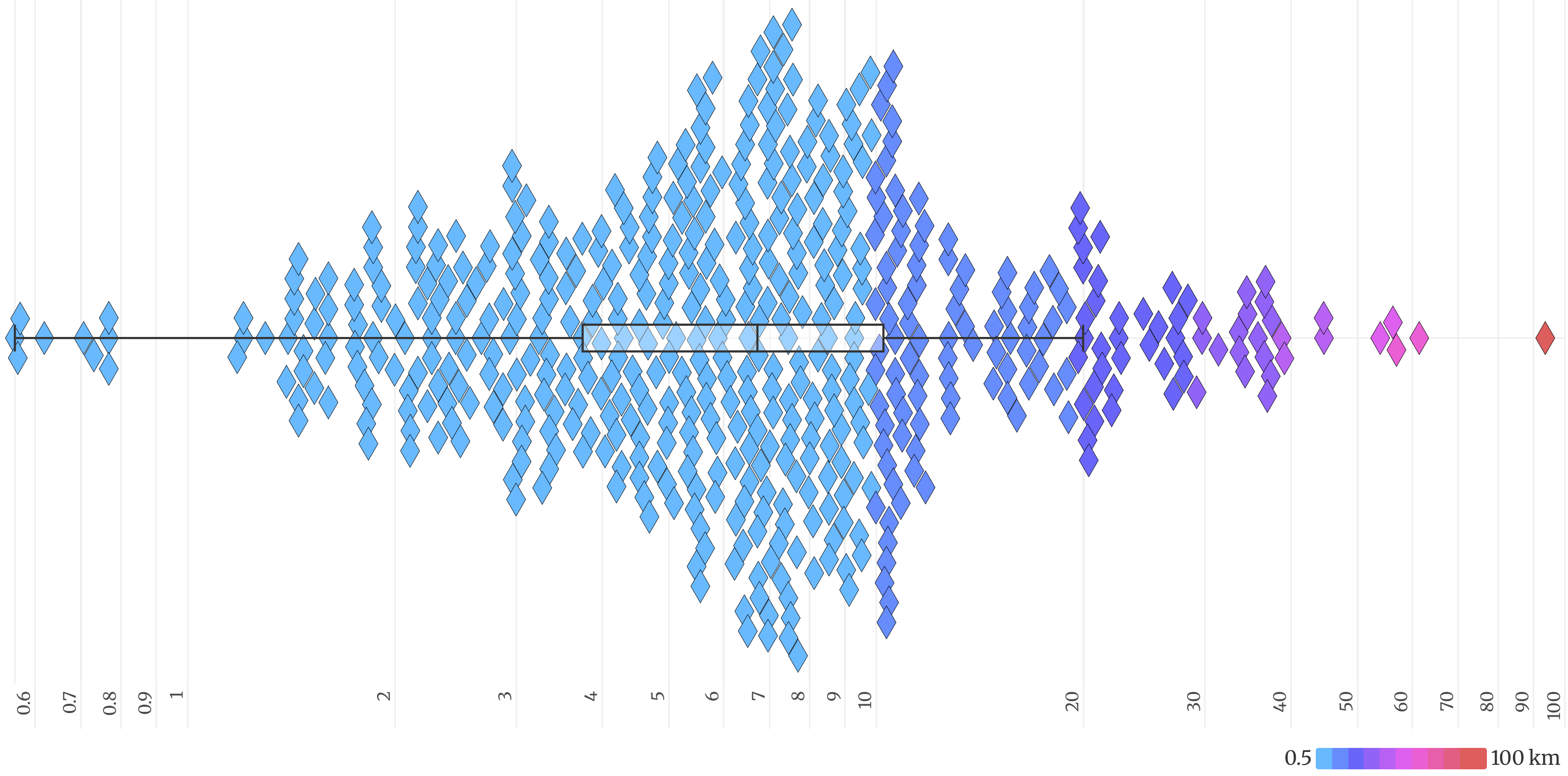}
    \caption{Combined beeswarm and boxplot of commuting distances (in kilometres, logarithmic scale) for survey respondents based on the routing analysis for the driving mode, averaged across campus-bound and home-bound trips. \it{Note: The respondents with missing stop information and zero commuting distances are excluded from the analysis.}}
    \label{fig:commute-distances}
\end{figure}

\paragraph{Commuting Distances:}
\autoref{fig:commute-distances} shows the commuting distances (in kilometres, logarithmic scale) for survey respondents based on the routing analysis for the driving mode, averaged across campus-bound and home-bound trips.
Driving was selected for this analysis as private car, carpooling, and shared commutes are collectively the most used mobility modes by respondents (see \autoref{fig:commute-modes-transfers-summer} and \autoref{fig:commute-modes-transfers-winter}).
In addition, driving allows for a more direct comparison of distances, as it is less affected by the availability of public transit routes or cycling infrastructure.

\autoref{fig:commute-distances} combines a beeswarm plot with a boxplot overlay, where the x-axis represents the average distance in kilometres and the y-axis shows the number of respondents.
The beeswarm plot displays the individual commuting distances, while the boxplot provides a summary of the distribution, including the median, the interquartile range (IQR) and outliers beyond the 1.5*IQR threshold.
Note that respondents with missing stop information and zero commuting distances are excluded from the analysis, leaving 547 (95.4\%) of all respondents in the analysis.

The analysis reveals commuting distances ranging from 561 m to 93.62 km, with a mean distance of 9.09 km and a standard deviation of 9.37 km. The median distance is 6.72 km, and the IQR ranges from 3.74 km to 10.24 km.
The distribution of commuting distances is right-skewed, with a few respondents having significantly longer commutes, as indicated by the outliers in the boxplot beyond the cluster of points around the 20 km mark.

\begin{figure}[tb!]
    \centering \includegraphics[width=.99\textwidth]{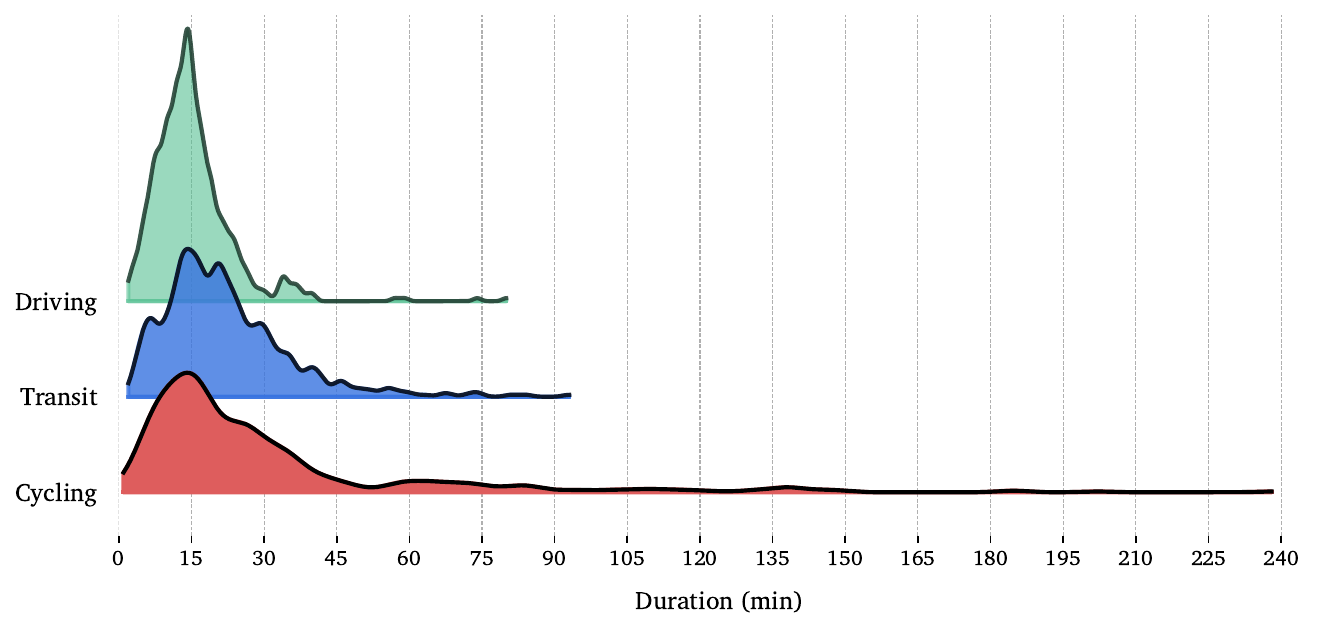}
    \caption{Joyplot of travel durations for different mobility modes, averaged across campus-bound and home-bound trips. \it{Note: The respondents with missing stop information, zero commuting distances, and incomplete routing data for public transit are excluded from the analysis.}}
    \label{fig:commute-times-modes}
\end{figure}

\paragraph{Travel Durations:}
\autoref{fig:commute-times-modes} shows the travel durations for the different mobility modes used by the survey respondents, averaged between campus-bound and home-bound trips.
The data is derived from the routing analysis for April 23, 2025 and exclude respondents with missing stop information, zero commuting distances, and incomplete routing data for public transit. This leaves 544 (94.9\%) of all respondents in the analysis.

\autoref{fig:commute-times-modes} uses a joyplot to compare the travel duration distributions for different mobility modes, with the x-axis representing the average travel duration in minutes and the y-axis showing the density of respondents for each mode.
\autoref{tab:commute-times-stats} provides the corresponding summary statistics, including the higher moments such as standard deviation (SD), skewness, and kurtosis, which indicate the shape of the distributions.
The upper section of the table shows the raw travel times for each mobility mode, while the lower section shows the differences in travel times between other modes and public transit as a baseline.

\begin{table}[width=.85\linewidth,pos=tb!]
    \renewcommand{\arraystretch}{1.15}
    \caption{Summary statistics for travel durations across different mobility modes. The ``diff'' tag indicates the difference between other modes and public transit. \it{Note: Respondents with missing stop information, zero commuting distances, and incomplete routing data for public transit are excluded from the analysis.}}
    \label{tab:commute-times-stats}
    \begin{tabular*}{\tblwidth}{@{\extracolsep\fill} l|ccccccccc@{}}
        \toprule
        Mode & Mean & SD & Minimum & 25\% & 50\% & 75\% & Maximum & Skewness & Kurtosis \\
        \midrule
        Driving & 15.17 & 8.28 & 2.0 & 10.0 & 14.0 & 18.0 & 80.0 & 2.56 & 13.5 \\
        Transit & 21.88 & 13.1 & 2.0 & 13.0 & 20.0 & 28.0 & 93.0 & 1.6 & 4.2 \\
        Cycling & 29.94 & 31.2 & 1.0 & 12.75 & 20.0 & 33.0 & 238.0 & 2.76 & 9.46 \\
        \midrule
        diff.Driving & -6.71 & 6.85 & -40.0 & -10.0 & -5.0 & -1.0 & 8.0 & -1.21 & 2.04 \\
        diff.Cycling & 8.06 & 20.84 & -15.0 & -2.0 & 1.0 & 8.25 & 154.0 & 3.32 & 13.62 \\
        \bottomrule
    \end{tabular*}
\end{table}

The analysis of the joyplot distributions and summary statistics reveals that the distributions of raw travel times vary substantially across mobility modes.
Driving exhibits the shortest typical travel times, with a mean of 15.17 minutes and a median of 14 minutes, and an IQR of 10 to 18 minutes. Transit travel times are longer on average, with a mean of 21.88 minutes and a median of 20 minutes, and a wider IQR of 13 to 28 minutes. Cycling has the longest central tendency, with a mean of 29.94 minutes and a median of 31 minutes, and an IQR of 20 to 33 minutes.

All three modes display right-skewed distributions, indicating the presence of respondents with substantially longer commute times than the majority. The skewness is most pronounced for cycling (2.76) and driving (2.56), and to a lesser extent for transit (1.60). Kurtosis values are high for driving (13.50) and cycling (9.46), signalling sharply peaked distributions with heavy tails, while transit (4.20) is less extreme but still heavier-tailed than a normal distribution.

\begin{figure}[tb!]
    \centering \includegraphics[width=.99\textwidth]{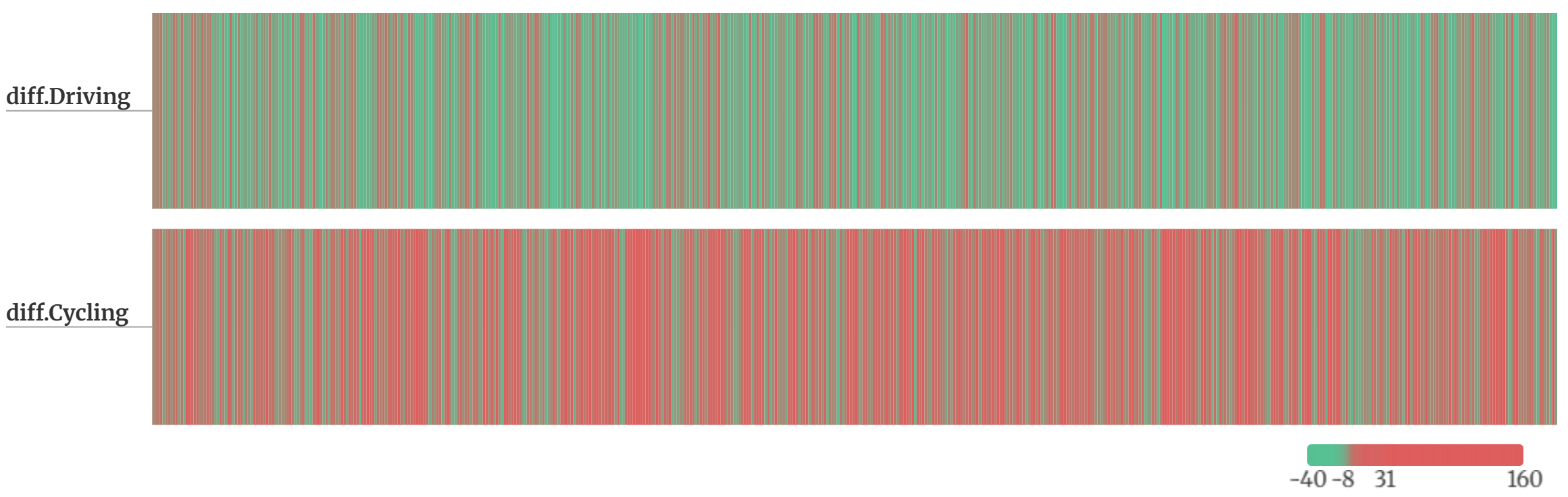}
    \caption{Numeric heatmap of travel time differences between mobility modes and public transit. The x-axis represents individual respondents, while the y-axis shows relative duration differences. \it{Note: The respondents with missing stop information, zero commuting distances, and incomplete routing data for public transit are excluded from the analysis.}}
    \label{fig:commute-times-diffs}
\end{figure}

The lower section of \autoref{tab:commute-times-stats} shows the differences in travel times between other mobility modes and public transit.
\autoref{fig:commute-times-diffs} uses a numeric heatmap to visualise these differences, with the x-axis representing individual respondents and the y-axis showing how their travel durations differ from public transit.
In the table and heatmap, negative values (green bars) indicate that the mode is faster than public transit, while positive values (red bars) indicate that the mode is slower than public transit.

Driving is generally faster than public transit, with a mean difference of $-6.71$ minutes and a median difference of $-5.0$ minutes. The negative skew ($-1.21$) reflects the cases where driving is much faster than transit, though the low kurtosis (2.04) indicates that these cases are not extreme outliers.
Cycling tends to be slower than transit, with a mean difference of $+8.06$ minutes but a median difference of only $+1.0$ minutes. The strong positive skew (3.32) and very high kurtosis (13.62) indicate that while most cycling trips are close in duration to transit, there are occasional and extremely long cycling commutes.

These differences highlight that, for most respondents, driving offers consistent time savings relative to public transit, whereas cycling times present more variability. The advantages in travel time of driving and cycling over public transit are primarily due to the directness of the routes and the absence of transfers and waiting times.
In addition, the duration of public transit travel is influenced by route availability, fixed paths designed to cover larger areas, and frequent stops for pickups and dropoffs, which can result in longer travel times compared to driving or cycling.

\begin{figure}[tb!]
    \centering \includegraphics[width=.99\textwidth]{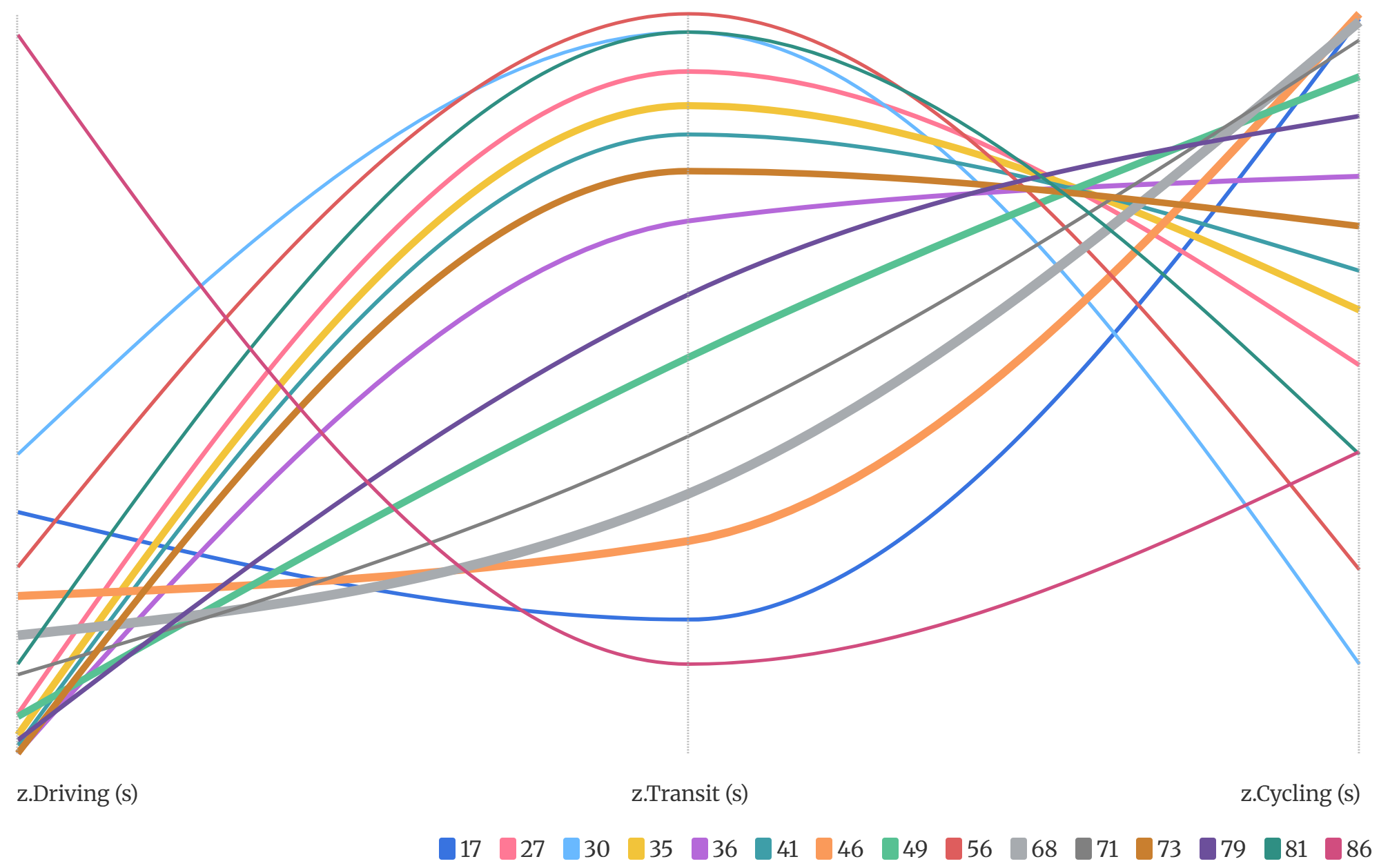}
    \caption{Slope chart of $z$-scores for 15 sampled ``most affected'' respondents, comparing travel durations across mobility modes. Line thickness reflects raw public transit travel duration. \it{Note: The respondents with missing stop information, zero commuting distances, and incomplete routing data for public transit are excluded from the analysis.}}
    \label{fig:commute-times-affected}
\end{figure}

\paragraph{Affected Individuals:}
The relative travel duration differences between mobility modes and public transit shown in \autoref{fig:commute-times-diffs} can be further explored to better understand the individual commuting realities.
\autoref{fig:commute-times-affected} zooms in on a sample of respondents drawn from those among the 544 respondents potentially ``most affected'' by parking reductions, as defined by the following criteria:

\begin{itemize} \itemsep=0pt
    \item Respondents who indicated that reduced parking would definitely or possibly affect their work schedule as shown in \autoref{fig:parking-commute-time-transfers}.
    \item Respondents who anticipated changes in their work schedule, as indicated by the difference between their current and preferred start and end times, as shown in \autoref{fig:work-hours-change-combined}.
\end{itemize}

These criteria identified 97 respondents, 16.9\% of the entire sample of 573. Of these, 90 (15.7\%) are among the 544 respondents with complete routing data for the three modes (driving, public transit, and cycling).
To ensure that the visualisation reflects the same proportion of affected respondents as in the complete dataset, a sample of 15 respondents was drawn, calculated as \mt{round(97/573*90)} to preserve this ratio while keeping the figure legible.
To compare \emph{within-respondent} differences between modes, \emph{row-wise} standardisation was applied to the travel durations for each respondent $i$ and mode $m$ such that:
$$ z_{i,m} = \big(t_{i,m} - \bar t_i\big)\big/ s_i $$

where $\bar t_i$ and $s_i$ are the mean and standard deviation of $(t_{i,\mathrm{Cycling}}, t_{i,\mathrm{Transit}}, t_{i,\mathrm{Driving}})$.
This yields $z$-scores with mean $0$ and standard deviation $1$ \emph{per respondent}. Positive $z$ indicates a mode slower than that respondent’s average; negative $z$ indicates faster. Magnitudes are comparable across respondents in within-person SD units, but the absolute times and variance between respondents are intentionally removed.

\autoref{fig:commute-times-affected} presents a slope chart illustrating the $z$-scores for 15 sampled ``most affected'' respondents. The x-axis represents mobility modes, while the y-axis shows $z$-scored travel durations. Each respondent is depicted by a coloured line connecting their $z$-scores across the three modes, the line thickness reflecting the raw duration of public transit travel. The chart reveals the following trends:

\begin{itemize}
    \item For short commutes (thin lines), driving and cycling are significantly faster than public transit (e.g., respondents 27 and 41). For very short commutes, cycling surpasses driving due to its direct routes and immunity to traffic delays (e.g., respondents 30 and 56).
    \item For medium commutes (medium lines), public transit becomes competitive with cycling, leveraging its speed advantage over cycling's directness. In some cases, public transit outpaces driving, probably due to dedicated lanes and priority signals (e.g., respondents 17 and 86). However, cycling can still outperform public transit in certain scenarios due to its direct routes and lack of waiting times (e.g., respondents 35 and 73).
    \item For long commutes (thick lines), driving and public transit converge as delays from transfers and waiting times in public transit become less significant relative to the overall travel time (e.g., respondents 46 and 68). Driving remains faster, but the difference narrows. Unsurprisingly, long commutes show the highest cycling $z$-scores, indicating cycling's limited feasibility for such distances.
\end{itemize}

\subsubsection{Congestion Analysis: Driving versus Cycling}
To complement the previous analyses, this section performs a rudimentary congestion analysis that compared the home--campus traffic flow for driving and cycling mobility modes, focussing on the main campus area at Gløshaugen.
The analysis is based on the same routing data as in \autoref{sec:results-distances-durations}, in this case using the route segments returned by the TravelTime API. These segments list the individual parts of the route in sequential order, each of which includes various details, including the coordinates of the route segment.

\begin{figure}[tb!]
    \centering
    \begin{subfigure}[b]{.99\textwidth}
        \centering \includegraphics[width=\textwidth]{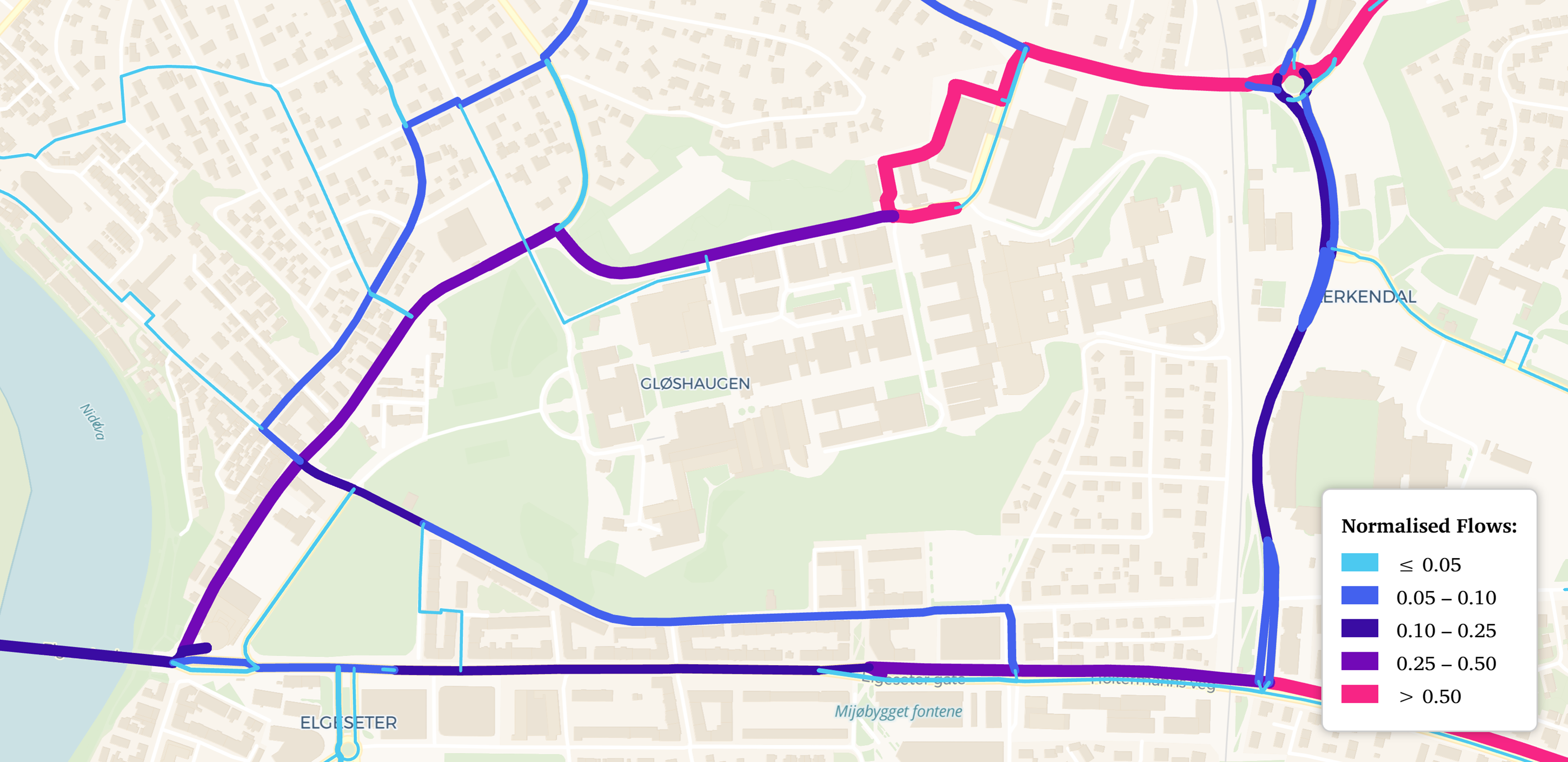}
        \caption{Driving} \label{fig:route-segments-driving}
    \end{subfigure}
    \begin{subfigure}[b]{.99\textwidth}
        \centering \includegraphics[width=\textwidth]{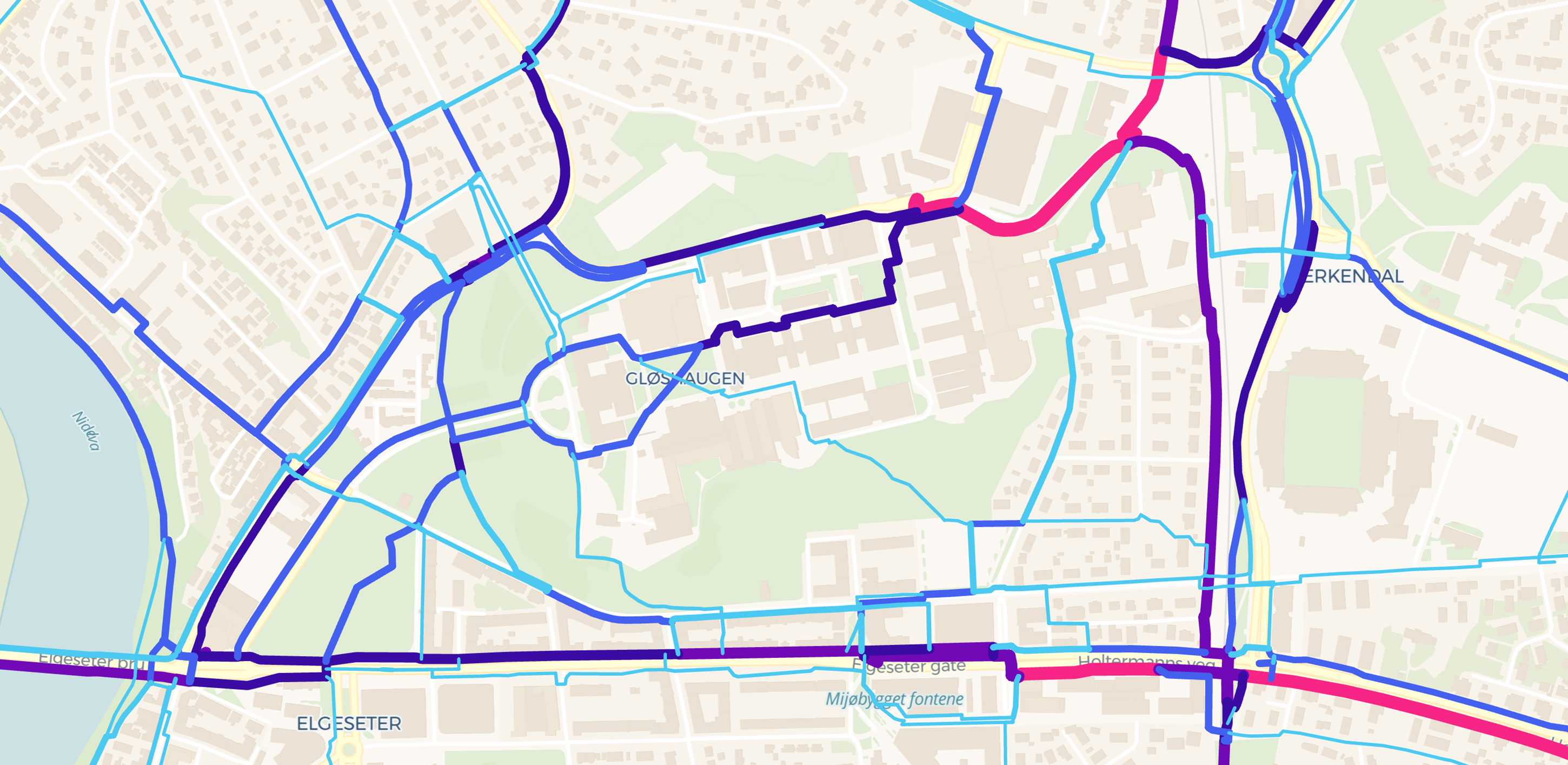}
        \caption{Cycling} \label{fig:route-segments-cycling}
    \end{subfigure}
    \caption{Map of the main campus area at Gløshaugen, showing congestion analysis results for driving and cycling. The segments are coloured according to their normalised traffic counts, with the same colour scheme used in both subfigures. \it{Note: The map is rotated to align with the major access routes along ``Elgeseter Gate'', with north pointing to the left side of the figure. The respondents with missing stop information are excluded from the analysis.}}
    \label{fig:route-segments-comparison}
\end{figure}

The analysis focusses on identifying potential congestion and bottlenecks in traffic flow if all respondents were to use the same mode of transport, either driving or cycling, for their home--campus commutes. It is based on data from 564 (98.4\%) respondents who provided complete routing information for both modes, excluding cases with missing stop data.
For each mode, route segments are aggregated by their coordinates to create a set of unique segments, with counts representing the number of respondents using each segment. These counts are min-max normalised to a range of 0 to 1, enabling direct comparisons of relative segment usage across mobility modes.

\autoref{fig:route-segments-comparison} presents the results of the congestion analysis, with \autoref{fig:route-segments-driving} and \autoref{fig:route-segments-cycling} showing the segments for driving and cycling, respectively. The route segments are displayed on a map of the main campus area in Gløshaugen, coloured according to their normalised counts as indicated in the legend. The same colour scheme and scale are applied to both subfigures for consistency.
Note that the map is rotated to align with the major access routes along ``Elgeseter Gate'', with the north pointing to the left side of the figure.

The analysis reveals that the traffic flow for driving is less evenly distributed in the main campus area compared to cycling. The roads surrounding the main campus see consistent usage, which tapers off quickly further away. The segments with high traffic are located mainly to the south and east of the main campus area, along ``Strindvegen (Øst)'' and ``Holtermanns Veg''.

In contrast, cycling routes are more evenly distributed throughout the campus area, due to the ability to use paths and lanes that are not accessible to cars. There is more traffic on roads farther away from the main campus, showing the greater flexibility and directness of the cycling routes compared to driving. The segments with high traffic are still located to the south and east of the main campus area. However, driving traffic along ``Strindvegen (Øst)'' now shifts to the cycling paths that feed into ``Høgskoleringen''.

\subsubsection{Summary and Implications}
The mobility analysis reveals clear seasonal patterns in commuting behaviour, with cycling usage dropping significantly in winter as public transit and driving increase. Although many respondents view public transit as a feasible alternative to private car use, its competitiveness varies with distance. For shorter commutes, driving and cycling offer better travel times, while for longer distances, public transit becomes more viable despite some identified connection gaps that require improvement.

The combination of directed flows and congestion analysis highlights potential bottlenecks around the main campus, particularly along ``Strindvegen'' and ``Holtermanns Veg''. These areas, along with the main transit stops in ``Gløshaugen'', ``Hesthagen'', and ``Høgskoleringen'', may require capacity considerations to accommodate the evolving campus population and mobility patterns.

The analysis shows that reducing parking access produces measurable secondary effects: longer commute times, increased stress and lower perceived productivity. In systemic terms, these are negative feedback signals that should trigger policy recalibration. Incorporating such feedback in real time would transform static policy evaluation into an adaptive governance model.

\vspace{\baselineskip}

\section{From Isolated Policy to Systemic Governance: A Cybernetic Perspective}
\label{sec:cybernetics}

While the preceding results describe behavioural and perceptual effects of NTNU's parking policy, these findings also reveal something deeper about how sustainability policies operate as systems. The data expose multiple layers of feedback between institutional design, human behaviour and the urban environment. This section outlines how the NTNU case could exemplify an adaptive feedback architecture for future research, linking policy design, data and learning.

Modern sustainability challenges cannot be solved by adjusting single parameters such as parking fees or commuting incentives. They require policies that learn, designed as adaptive systems capable of sensing feedback, interpreting it, and modifying themselves accordingly. This notion is deeply rooted in cybernetics, the science of steering complex systems through continuous information feedback \citep{Wiener2019coc,Beer1995bot,Meadows2008tis}. Drawing from engineering cybernetics, such systems must be controllable, observable, and adaptive to sustain stability under changing conditions. This perspective connects with broader work on systemic intervention and dynamic modelling \citep{Midgley2000si,Sterman2009bds}, as well as with recent transport studies emphasising behavioural complexity, demand-side solutions, and context-specific challenges in sustainable urban mobility \citep{Banister2008tsm,Creutzig2022dss,Canitez2019pts}.

A cybernetic governance model assumes that policy is never static, but operates through iterative loops of policy $\rightarrow$ behaviour $\rightarrow$ data $\rightarrow$ policy revision. The NTNU mobility case illustrates what happens when these loops are incomplete. Emerging research on multilevel policy mixes highlights similar challenges, where national funding biases and limited cross-level cooperation hamper sustainable urban mobility transitions \citep{Liu2024imp,Geels2012ast}. Furthermore, studies of mobility exclusion demonstrate that infrastructure-focused approaches often fail to address the systemic nature of transport disadvantage, requiring more holistic governance approaches that integrate diverse perspectives \citep{OliveiraSoares2025biu}.

Traditional transport analyses optimise for direct financial or environmental indicators (cost, emissions, parking supply) but rarely for the systemic consequences on human well-being, organisational productivity and social cohesion. Our mixed-methods results show how employees respond adaptively to top-down restrictions: they modify commuting patterns, work hours and, in some cases, employment intentions. These are not mere side-effects; they are feedback signals from the system, indicating that the policy is generating new conditions that demand recalibration. From a systems-architectural viewpoint, each data source in this study represents one layer of feedback.

Survey responses reveal subjective experience and behavioural intent; mobility data capture actual flows; and demographic variables expose structural constraints such as family, housing and distance. When these layers are integrated, they form a multi-level feedback architecture capable of showing how policy propagates through the social and physical fabric of the university. This approach resonates with research on citizen involvement in mobility planning, which emphasises the need for multi-level participation frameworks that connect strategic and tactical decision-making \citep{Ibeas2011cii}.

In this light, NTNU can be viewed as a microcosm of national governance, a bounded sociotechnical system where policies can be prototyped and evaluated in real time. Embedding cybernetic principles into campus management would mean treating every policy as an experiment with measurable loops: Sensing (data), interpretation (analysis), decision (revision), and action (implementation). This aligns with findings from studies of transport poverty, which show that generic policy measures often prove ineffective without targeted interventions informed by detailed understanding of local vulnerabilities \citep{DallaLonga2025sit}. A practical embodiment of this idea could be an adaptive policy dashboard linking survey feedback, commuting data, and productivity indicators to guide iterative improvements.

\begin{figure}
    \centering
    \includegraphics[width=.99\textwidth]{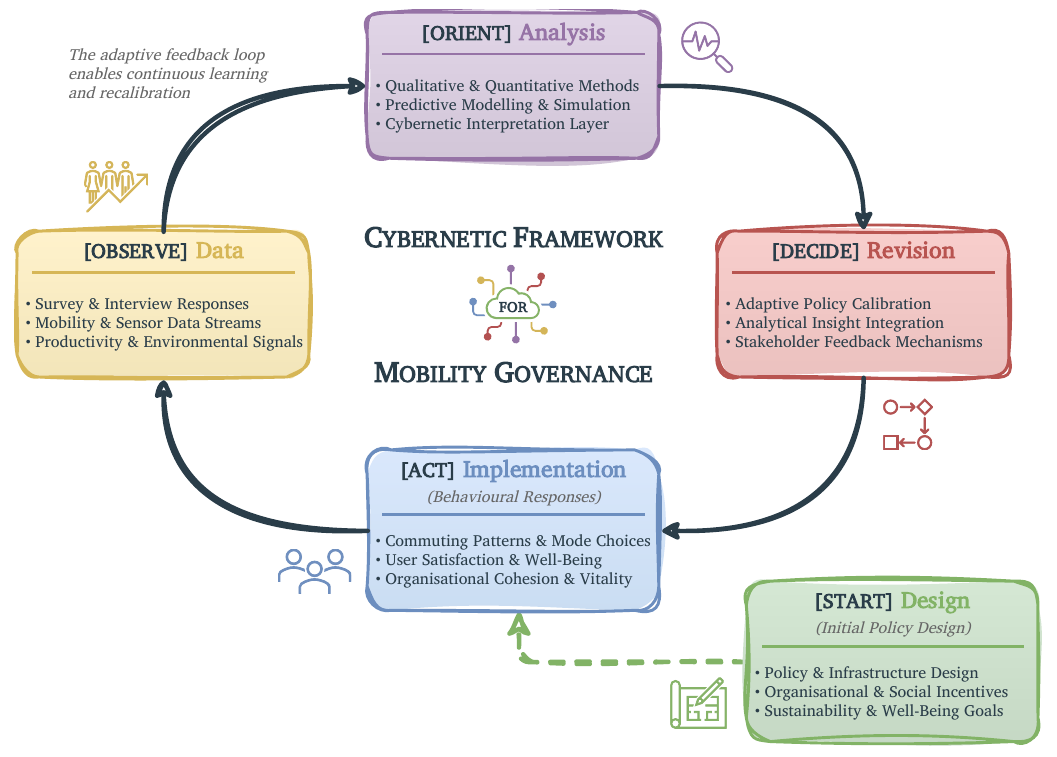}
    \caption{Cybernetic feedback architecture for adaptive mobility governance. The framework operationalises policy as a learning system where design intentions elicit behavioural responses, which are sensed as data, interpreted analytically, and used to recalibrate subsequent actions. \it{Note: The four-stage loop draws on principles from engineering cybernetics and is analogous to an OODA (Observe-Orient-Decide-Act) loop.}}
    \label{fig:cybernetic-feedback-architecture}
\end{figure}

Such a systemic model differs fundamentally from traditional policy evaluation, as shown in \autoref{tab:governance-comparison}. Conventional approaches measure outputs (for example, how many people stopped driving to work), while cybernetic evaluation measures system performance: ``how effectively does the policy maintain the balance between sustainability, accessibility, and organisational vitality?'' This shift in perspective embodies the core tenet of engineering cybernetics by replacing static success criteria with dynamic equilibrium: a state in which the system continuously self-corrects to maintain function without eroding human capacity. In line with this view, \autoref{fig:cybernetic-feedback-architecture} illustrates a cybernetic framework for adaptive governance that operationalises policy as a learning system.

\begin{table}[width=.99\linewidth,pos=tb!]
    \renewcommand{\arraystretch}{1.15}
    \caption{Comparison of traditional policy analysis and systemic cybernetic approach to governance.}
    \label{tab:governance-comparison}
    \begin{tabular*}{\tblwidth}{@{\extracolsep\fill} lll@{} }
        \toprule
        Dimension & Traditional Policy Analysis & Systemic Cybernetic Approach \\
        \midrule
        Focus & Isolated targets (reduce car use) & Interconnected mechanisms (mobility, well-being, productivity) \\
        Time Scale & Snapshot evaluation & Continuous feedback and adaptation \\
        Data Use & Descriptive and retrospective & Predictive and integrative \\
        Governance Logic & Control and compliance & Learning and self-regulation \\
        Outcome & Efficiency & System resilience and long-term sustainability \\
        \bottomrule
    \end{tabular*}
\end{table}

The architecture begins with a \it{design} phase that establishes rules and incentives, followed by a four-stage adaptive feedback loop that conceptually extends Boyd's OODA framework \citep{Boyd2018ado,Osinga2007ssa} from strategic decision-making for continuous policy learning. First, \it{implementation} generates behavioural responses across mobility patterns, satisfaction, and organisational dynamics. Second, systematic \it{data} collection captures these responses through surveys, sensors, and performance indicators. Third, \it{analytical} interpretation detects deviations from desired system performance by integrating system dynamics and cybernetic interpretation layers. Finally, policy \it{revision} integrates this feedback to recalibrate subsequent interventions. This closed-loop structure transforms linear policy implementation into an adaptive system where each iteration strengthens the alignment between sustainability goals and human experience by integrating behavioural adaptation as essential system feedback.

When applied to the NTNU case, this cybernetic lens reframes the empirical findings. The measured increase in commuting time and stress is not simply a cost; it represents system feedback indicating operation outside an optimal performance range. In cybernetic terms, these signals should trigger negative feedback (leading to policy revision) rather than being ignored as statistical noise. Likewise, the observed heterogeneity across distance, family status, and season points to interaction effects that policy design must explicitly model rather than average away. Research on effective policy packages for sustainable commuting emphasises similar points, showing that successful interventions must address all elements of practice and wider networks of practices \citep{Albers2025dep}.

Ultimately, the NTNU case provides a prototype for how universities and municipalities can develop learning-based governance architectures. Instead of designing one-off interventions, institutions can cultivate positive feedback loops that amplify collaboration, innovation, and collective intelligence, while dampening the negative effects of stress, inefficiency, and disengagement. The goal is not merely sustainable mobility but sustainable systems intelligence---the ability of an organisation to sense itself, learn and adapt in real time. This approach must account for the uncertainties inherent in sustainability transitions, requiring exploratory approaches that can identify viable pathways under uncertain futures \citep{Moallemi2019cwu}.

\section{Conclusion}
\label{sec:conclusion}
This study demonstrates how sophisticated mixed-methods approaches can bridge the gap between the formation and implementation of sustainable mobility policies through enhanced analytical capabilities. Through the analysis of NTNU's campus consolidation, we reveal both the complexity of mobility transitions and the potential for data-driven solutions to address implementation challenges while maintaining sustainability goals. The integration of qualitative survey insights with quantitative big mobility data provides a comprehensive understanding of mobility patterns, constraints, and policy impacts that can inform more effective and inclusive policy design. The findings also demonstrate how such analytical frameworks can inform adaptive governance approaches that treat data as feedback for continuous policy learning and improvement.

\subsection{Summary of Key Findings}
\vspace{0.5\baselineskip}

\paragraph{Mixed-Methods Framework and Enhanced Analytical Capabilities}
This study developed a comprehensive mixed-methods framework that combined qualitative survey data (n=573) with quantitative big mobility data analysis to understand spatiotemporal mobility patterns, constraints, and policy impacts. The framework integrated three analytical components: (1) qualitative survey analysis that reveals lived mobility experiences and barriers, (2) quantitative analysis of public transit use and crowd movement patterns that capture actual flows, and (3) comparative mobility analysis that examines travel durations, distances and congestion patterns across driving, cycling and public transit modes. This integration demonstrates how analytical frameworks can inform adaptive governance approaches that interpret outcomes as feedback for policy learning.

The spatiotemporal understanding developed through this framework revealed pronounced temporal patterns with morning peaks (08:00--09:00) and afternoon peaks (15:00--17:00), significant seasonal variations in mobility mode usage (cycling decreasing 58.3\% from summer to winter), and distance-dependent competitiveness of different transport modes. The key identified needs include flexible mobility solutions for the 59.3\% of respondents with children, accessible options for those with health conditions, improved frequency and directness of public transit, improved cycling infrastructure with safe and secure indoor parking and family-friendly scheduling accommodations.

\paragraph{Policy Formation and Implementation Gaps as System Signals}
Several critical gaps between policy formation and implementation were discovered, functioning as negative feedback signals indicating that the system is operating outside its optimal range. The analysis revealed that although 73.3\% of the respondents consider public transit a feasible alternative to driving, the actual usage remains only 22.9\%, indicating a substantial implementation gap. In cybernetic terms, these signals should trigger policy adjustment rather than being ignored as statistical noise.

The complexity of family logistics appears to have been underestimated, with 86.0\% of the respondents identifying increased travel duration as a primary difficulty and 42.1\% anticipating increased mobility costs. The current policy framework inadequately addresses diverse life situations, particularly for the 15.7\% who would consider leaving NTNU due to parking reductions, representing potential organisational consequences of systemic stress. The policy lacks sufficient consideration of health and accessibility needs, seasonal mobility variations, and the practical realities of multimodal trip chaining for daily activities. Policies that optimise for infrastructure cost may undermine human-system performance by increasing stress and commute time.

\paragraph{Behavioural Changes and Congestion Risks}
The study identified several potential behaviour changes and congestion scenarios. Among the 286 respondents who anticipate schedule changes, 64.9\% expect later work starts and 67.0\% earlier finishes, potentially alleviating peak hour pressure but reducing collaboration time. In addition, 85.7\% of the 77 respondents considering changes in their work patterns expect to reduce their presence on campus, shifting to remote work.

The congestion analysis revealed potential bottlenecks along key access routes to campus, with driving routes showing less even distribution compared to cycling. The major public transit stops already handle substantial volumes (more than 6,000 passengers during peak hours), indicating capacity limitations that could worsen with increasing demand from campus consolidation.

\paragraph{Broader Implications and Future Research Directions}
The disconnect between the formation and implementation of sustainable mobility policies observed at NTNU reflects broader challenges facing policymakers worldwide as they pursue mobility transitions. The mixed-methods framework developed in this study offers a transferable cybernetic approach to understanding these dynamics in diverse urban contexts, from university campuses to city-wide mobility planning initiatives.

Future research directions include extending the methodology to analyse big data from multimodal trip chains and additional transport modes, in addition to employing more advanced analytical and modelling techniques. Research on multimodal mobility hubs demonstrates the potential for decision support tools that optimise accessibility to both workplaces and places of private need \citep{Frank2021ira}. Longitudinal studies could track behavioural changes over time as policies are implemented, providing insight into adaptation processes and long-term impacts on sustainable mobility transitions.

The results show the potential for NTNU's campus to serve as a living laboratory for national mobility governance, where policies can be tested and recalibrated in real time. Such an approach would mark a transition from static policy evaluation to cybernetic, evidence-based learning—a necessary step for sustainable governance in complex societies. Integrating employee perception data with objective mobility metrics creates a feedback architecture for adaptive decision-making.

\subsection{Policy Recommendations}
The study findings inform recommendations at two levels: specific interventions for the NTNU case study and broader methodological approaches to apply mixed-methods frameworks to urban mobility policy.

\paragraph{Case-Specific Recommendations for NTNU}
The analysis of NTNU's campus consolidation reveals the need for targeted infrastructure and policy interventions. These recommendations address the immediate challenges identified in our study while providing a foundation for sustainable mobility transitions.

Infrastructure improvements include secure indoor bicycle parking with 2,462 planned spaces in six projects, improved public transit frequency during peak hours (08:00--09:00 and 15:00--17:00), improved cycling path winter maintenance, and targeted capacity increases at key transit stops (Gløshaugen, Hesthagen, Høgskoleringen).

Policy measures should include distance-based parking prioritisation, health-condition parking permits, flexible remote work policies, discounted public transit passes (preferred by 60.7\% of the respondents), and family-friendly scheduling accommodations. Future considerations should include electric vehicle charging infrastructure as part of a comprehensive sustainable transport ecosystem.

\paragraph{Methodological Recommendations for Cybernetic Mobility Governance}
This study demonstrates the value of combining qualitative and quantitative approaches for the development of adaptive governance architectures in mobility policy. The following recommendations outline how stakeholders can integrate survey insights with big data analysis to create learning-based feedback systems for more effective urban mobility interventions:

\begin{itemize} \itemsep=0pt
    \item \textbf{Municipal Authorities}: Deploy digital twin platforms integrating community feedback with real-time mobility flows for proactive infrastructure planning and participatory policy development.
    \item \textbf{Transport Authorities}: Implement adaptive management systems combining ridership analytics with user experience surveys for route optimisation and targeted service adjustments.
    \item \textbf{Technology Providers}: Develop interoperable platforms merging qualitative feedback with quantitative mobility data, enabling real-time policy impact assessment and scenario simulations.
    \item \textbf{Research Institutions}: Use combined qualitative and quantitative analysis to simulate policy interventions and assess impacts before implementation, anticipating negative externalities through a comprehensive stakeholder analysis.
    \item \textbf{Funding Agencies}: Prioritise research bridging technical innovation with social equity analysis, supporting mixed-methods tools for rapid policy evaluation across diverse urban contexts.
\end{itemize}

The findings of this study point to a fundamental transformation in how sustainability policies can be conceived and implemented: moving beyond traditional approaches that measure isolated outcomes to adaptive governance architectures that interpret policy effects as feedback signals within complex socio-technical systems. In the long term, the transition from descriptive analytics to reflexive, cybernetic governance will be essential for cities and universities seeking to align sustainability objectives with human well-being.

\end{document}